\documentclass[12pt,preprint]{aastex}
\usepackage{epsf}
\usepackage{graphicx}
\usepackage{natbib}
\usepackage{float}

\tabletypesize{\scriptsize}

\newcommand{\hd}{{HD 181327}}
\def\gappeq{\mathrel{ \rlap{\raise.5ex\hbox{$>$}}
                      {\lower.5ex\hbox{$\sim$}}  } }

\begin{document}
\slugcomment{Accepted for Publication in ApJ}
\shorttitle{Asymmetries in the HD 181327 Debris Disk}
\shortauthors{}

\title{Revealing Asymmetries in the HD 181327 Debris Disk: A Recent Massive Collision or ISM Warping}

\author{Christopher C. Stark\altaffilmark{1}, Glenn Schneider\altaffilmark{2}, Alycia J. Weinberger\altaffilmark{3}, John H. Debes\altaffilmark{4}, Carol A. Grady\altaffilmark{5}, Hannah Jang-Condell\altaffilmark{6}, Marc J. Kuchner\altaffilmark{7}}

\altaffiltext{1}{NASA Goddard Space Flight Center, Exoplanets \& Stellar Astrophysics Laboratory, Code 667, Greenbelt, MD 20771; christopher.c.stark@nasa.gov}
\altaffiltext{2}{Steward Observatory, The University of Arizona, Tucson, AZ 85721}
\altaffiltext{3}{Department of Terrestrial Magnetism, Carnegie Institution of Washington, 5241 Broad Branch Road NW, Washington, DC 20015}
\altaffiltext{4}{Space Telescope Science Institute, Baltimore, MD 21218}
\altaffiltext{5}{Eureka Scientific, 2452 Delmer, Suite 100, Oakland, CA 96002}
\altaffiltext{6}{Department of Physics \& Astronomy, University of Wyoming, Laramie, WY 82071}
\altaffiltext{7}{NASA Goddard Space Flight Center, Exoplanets \& Stellar Astrophysics Laboratory, Code 667, Greenbelt, MD 20771}

\begin{abstract}
New multi-roll coronagraphic images of the HD 181327 debris disk obtained using the Space Telescope Imaging Spectrograph (STIS) on board the Hubble Space Telescope (HST) reveal the debris ring in its entirety at high S/N and unprecedented spatial resolution.  We present and apply a new multi-roll image processing routine to identify and further remove quasi-static PSF-subtraction residuals and quantify systematic uncertainties.  We also use a new iterative image deprojection technique to constrain the true disk geometry and aggressively remove any surface brightness asymmetries that can be explained without invoking dust density enhancements/deficits.  The measured empirical scattering phase function for the disk is more forward scattering than previously thought and is not well-fit by a Henyey-Greenstein function.  The empirical scattering phase function varies with stellocentric distance, consistent with the expected radiation pressured-induced size segregation exterior to the belt.  Within the belt, the empirical scattering phase function contradicts unperturbed debris ring models, suggesting the presence of an unseen planet.  The radial profile of the flux density is degenerate with a radially-varying scattering phase function; therefore estimates of the ring's true width and edge slope may be highly uncertain.  We detect large scale asymmetries in the disk, consistent with either the recent catastrophic disruption of a body with mass $>1\%$ the mass of Pluto, or disk warping due to strong interactions with the interstellar medium (ISM).

\end{abstract}

\section{Introduction}
\label{intro}

Images of spatially resolved debris disks show a range of spectacular asymmetries including eccentric rings \citep[e.g.,][]{2005Natur.435.1067K, 2009AJ....137...53S, 2012AJ....144...45K, 2012ApJ...750L..21B}, warps and sub-structures \citep[e.g.,][]{2006AJ....131.3109G, 2005AJ....129.1008K}, and various other morphologies \citep[e.g.,][]{2007ApJ...671L.165H, 2007ApJ...661L..85K}.  Interpreting surface brightness asymmetries as dust density variations and explaining them via dynamic processes has been popular since coronagraphic images of the $\beta$ Pictoris disk revealed scattered light asymmetries \citep{1995AJ....110..794K}.  Recently, ALMA has begun to explore such patterns at unprecedented resolution at submillimter wavelengths \citep[e.g.,][]{2012ApJ...750L..21B}.

Many models have shown that exoplanets can create asymmetric dust distributions in debris disks via gravitational perturbations, potentially revealing the presence of otherwise undetectable planets.  Disk asymmetries created by planets could be the only way to detect true Neptune analogs orbiting nearby stars on reasonable timescales.  Direct images of exoplanet candidates associated with debris disks have begun to demonstrate both the potential and complexities of this concept for locating new planets and constraining their properties \citep[e.g.,][]{2006MNRAS.372L..14Q,2009ApJ...693..734C,2010Sci...329...57L,2013ApJ...775...56K}
 
Many other dynamical processes can also produce dust density asymmetries in debris disks.  Disks can interact with the interstellar medium \citep{1997ApJ...490..863A, 2009ApJ...702..318D, 2009ApJ...707.1098M, 2011MNRAS.416.1890M}.  In the solar system, collisions of asteroids can produce detectable trails of debris \citep{2010Natur.467..817J}.  In debris disks, recent collisions can potentially yield detectable arcs of debris \citep{2007A&A...461..537G,2013A&A...558A.121K,2014MNRAS.440.3757J}.

It remains crucial to pursue explanations for debris disk asymmetries other than density enhancements/deficits.  The observed scattered starlight from a disk is a complex combination of the disk's geometry, illumination, and stellocentric grain-size segregation, the dust grains' size-dependent scattering efficiency and scattering phase function, and line-of-sight projection effects in the case of inclined disks (e.g. ansal brightening for disks with large scale heights).  All of these effects are degenerate to some degree and, given our current understanding of dust grain scattering properties, are exceedingly difficult to disentangle.

In light of this, we interpret new multi-roll STIS coronagraphic images of the \hd\ debris disk by aggressively pursuing explanations for observed asymmetries \emph{other than} density enhancements.  We exploit a symmetry in the scattering phase function to search for deviations from a smooth disk.

\hd\ is an F5/6, $\sim12$ Myr old main sequence member of the $\beta$ Pic moving group, located at a distance of 51.8 pc \citep{2006ApJ...650..414S}.  \hd\ has a strong thermal IR excess ($L_{\rm IR}/L_{\star} = 0.25\%$) attributed to re-radiating circumstellar dust.  This debris disk was first detected with IRAS \citep{1998ApJ...497..330M} and subsequently resolved at $0.6\;\mu$m with the Advanced Camera for Surveys (ACS), $1.1\;\mu$m with the Near Infrared Camera and Multi-Object Spectrometer (NICMOS) on HST \citep{2006ApJ...650..414S}, $18.3\;\mu$m with Gemini South T-ReCS \citep{2008ApJ...689..539C}, and $3.2$ mm with the Australian Telescope Compact Array (ATCA) \citep{2012A&A...539A..17L}.  All resolved observations are consistent with a ring of dust with radius $\sim90$ AU and a cleared interior.  Most resolved images suggest clumpy asymmetries, but also have low S/R, so we refrain from incorporating those previous observations into our analysis.

In Section \ref{observations} we briefly describe our new STIS observations of \hd\ and present a new PSF residual removal routine.  For a more detailed description of the observations, see \citet{2014schneider_et_al}.  In Section \ref{results}, we detail our image deprojection techniques and ascertain the minimally-asymmetric face-on optical depth.  In Section \ref{discussion}, we interpret the observed scattering phase function and discuss several explanations for the observed asymmetries.

\section{Observations \& Data Reduction}
\label{observations}

As part of the HST GO 12228 (PI: G. Schneider) observation program, we observed HD 181327 in scattered light using the STIS coronagraph at 6 roll angles, each at 2 wedge-occulter positions (WedgeA0.6 and WedgeA1.0, which have occulting half-widths of $0.3$\arcsec and $0.5$\arcsec, respectively\footnote{See the HST STIS instrument handbook for a full description.}).  We observed HD 181327 over 3 orbits on 2011 May 20 and 3 orbits on 2011 July 10, and the ($B-V$ color-matched) PSF reference star HD 180134 at each wedge position and a single roll angle interleaved with the target observations.  Table \ref{observations_table} summarizes the observations.  The STIS 50CCD channel, equipped with the coronagraphic wedges used for our observations, has an image scale at the detector focal plane of $0.05077$ arcsec pixel$^{-1}$. Filters cannot be used with the coronagraphs, so the images are obtained with the full spectral response of the detector, i.e. a central bandpass of $0.5752$ $\mu$m and a FWHM of $0.433$ $\mu$m.  Multiple exposures in each observational configuration (a given roll angle and wedge position) are median-combined.   

\begin{deluxetable}{cccccc}
\tablewidth{0pt}
\footnotesize
\tablecaption{Observations \label{observations_table}}
\tablehead{
\colhead{Target} & \colhead{Orientation} & \colhead{Exposures} & \colhead{Total Exposure Time} & \colhead{Wedge} & \colhead{Date} \\
                               & \colhead{($^{\circ}$)} & & \colhead{(s)}\\
\vspace{-0.1in}
}
\startdata
HD 181327 & 222.738 & 8 & 175.2 & 0.6A & 20 May 2011 \\
 & 222.738   &  8 & 175.2  &   0.6A & 20 May 2011 \\
 & 222.738    & 4 & 87.6   &  0.6A & 20 May 2011 \\
 & 222.738   & 4 & 1708.0 &   1.0A & 20 May 2011 \\

  &                  242.738    & 8 & 175.2   &  0.6A & 20 May 2011\\
     &               242.738    & 8 &  175.2  &   0.6A & 20 May 2011\\
        &            242.738    & 4 &  87.6   &  0.6A & 20 May 2011\\
           &         242.738   & 4 & 1708.0 &    1.0A & 20 May 2011\\

HD 180134 &   243.219    &  8 & 95.2    &  0.6A  & 20 May 2011\\
  &  243.219    &  8 & 95.2     & 0.6A  & 20 May 2011\\
  &  243.219    &  8 & 95.2     & 0.6A  & 20 May 2011\\
  & 243.219   & 8 & 1656.0    &  1.0A & 20 May 2011\\

HD 181327    &                262.738    & 8 & 175.2  &   0.6A & 20 May 2011\\
       &             262.738    & 8 & 175.2  &   0.6A & 20 May 2011\\
          &          262.738    &  4 & 87.6  &   0.6A & 20 May 2011\\
             &       262.738   & 4 & 1708.0  &  1.0A & 20 May 2011\\

  &                  293.056    & 8 & 175.2  &   0.6A & 10 July 2011\\
     &               293.056    & 8 & 175.2  &   0.6A & 10 July 2011\\
        &            293.056    &  4 & 87.6  &   0.6A & 10 July 2011 \\
           &         293.056   & 4 & 1644.0 &  1.0A & 10 July 2011\\

  &                  313.556   &  8 & 175.2   &  0.6A & 10 July 2011\\
     &               313.556   &  8 & 175.2   &  0.6A & 10 July 2011\\
        &            313.556   &  4 &  87.6    &  0.6A & 10 July 2011\\
           &         313.556  & 4 & 1708.0 &    1.0A & 10 July 2011\\

HD 180134    & 314.232    &  8 & 95.2    &  0.6A & 10 July 2011\\
    & 314.232    &  8 & 95.2    &  0.6A & 10 July 2011\\
    & 314.232    &  8 & 95.2    &  0.6A & 10 July 2011\\
    & 314.232   & 8 & 1656.0  &   1.0A & 10 July 2011\\

HD 181327 &                   334.056   &  8 & 175.2   &   0.6A & 10 July 2011\\
    &                334.056   & 8 &  175.2  &    0.6A & 10 July 2011\\
       &             334.056    &  4 & 87.6   &   0.6A & 10 July 2011\\
          &          334.056   & 4 & 1708.0 &    1.0A & 10 July 2011\\

\enddata
\end{deluxetable}

All astrometric measurements of the HD 181327 debris ring are referenced to the position of the coronagraphically occulted star, as measured individually from twelve independent images using the diffraction spikes in each image to locate the star.  In each of the twelve images, the uncertainty in the target position is approximately 0.3 pixels (0.8 AU) in the Science Aperture Instrument (detector) Frame (SIAF) $x$ and $y$ directions.  In the twelve-image combination, the uncertainty is reduced to $\pm0.087$ pixels, or $\pm4.4$ mas (0.23 AU); this is comparable to the HST pointing stability (RMS 2-guide star fine-lock jitter).  The mean stellar position is used to anchor the inter- and intra-visit stellar location following target acquisition slews and to reference the WedgeA0.6 pointings to the WedgeA1.0 pointings.

\begin{figure}[H]
\begin{center}
\includegraphics[width=6in]{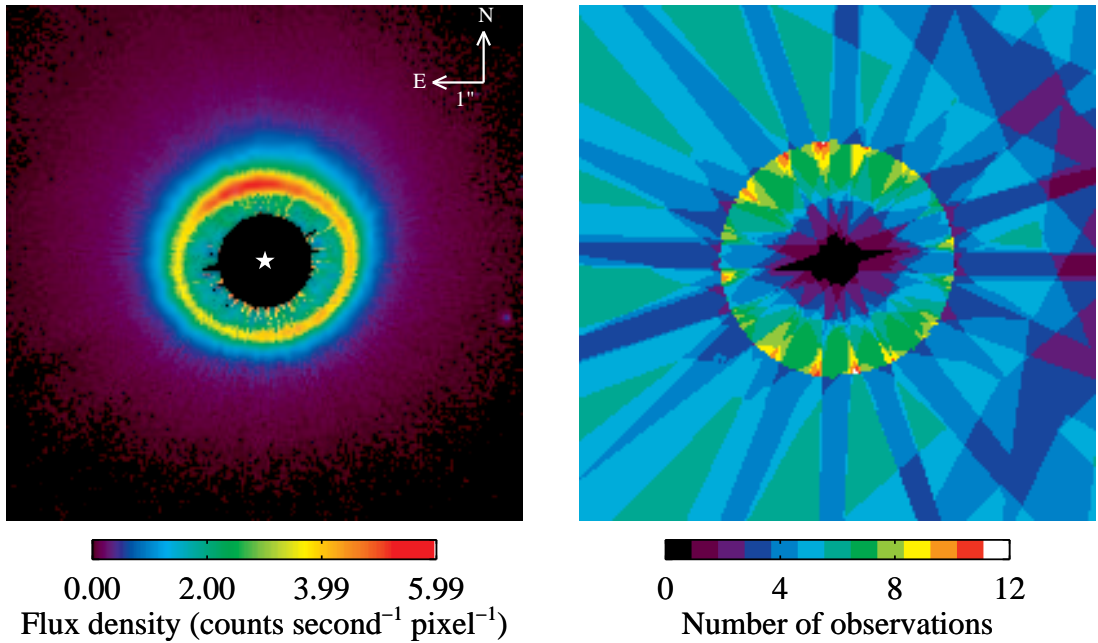}
\caption{\emph{Left:} Central 200$\times$200 pixels of the 12-image median-combined image for HD 181327 (pixel scale $=50.77$ mas $=2.63$ AU).  An artificial occulting spot with a radius of 18 pixels has been applied for illustrative purposes.  Stellar position is marked with a star.  \emph{Right:} Number of images used per pixel for median combination.  \label{twelve_image_median_fig}}
\end{center}
\end{figure}

All visit-level PSF subtractions are done in the SIAF, treating target and PSF-template positions and brightness as free parameters while iteratively minimizing PSF-subtraction residuals  following the procedure described by \citep{2009AJ....137...53S}. Each image is then rotated about the mean stellar position to a common north-up orientation.  We then manually build a mask specific to each observation, flagging those pixels that are obscured by the occulting wedge, corrupted by diffraction spikes, saturated or adversely affected by wedge-edge artifacts, or are beyond the FOV sub-array read out.  Details of this process and considerations for optimization are discussed in detail in \citet{2014schneider_et_al}.  Finally, we median-combine the 12 astrometrically co-registered images in the common celestial frame.  The left panel of Figure \ref{twelve_image_median_fig} shows the inner 200 $\times$ 200 pixels (pixel scale $= 50.77$ mas, or $2.63$ AU assuming a distance of $51.8$ pc) of our 12-image median.  Our multi-roll observation method reduces the mean inner working angle of the STIS coronagraph, reduces the influence of PSF artifacts, and improves S/R.  

The right panel of Figure \ref{twelve_image_median_fig} shows a map of the number of observations per pixel, $n_{\rm pix}$.  We mask off optical artifacts (occulting mask shadows, telescope diffraction spikes) that are rotationally invariant in the frame of the detector.  Thus, with multi-image masking, each output pixel in the final image is derived from different numbers of input images, so not all pixels in the final image are identically exposed.  The radial ``spokes" in this map are primarily a result of masking off the telescope's secondary mirror support structures at the 6 roll angles.  For the WedgeA0.6 observations, the images become photon starved at a radius of $\approx2.6\arcsec$, such that the S/N $\sim1$.  To avoid adding unnecessary noise to the outer disk, we masked the WedgeA0.6 images beyond $2.6\arcsec$.  As a result, the $n_{\rm pix}$ map exhibits a circular disk of radius $2.6\arcsec$, interior to which $n_{\rm pix}$ is larger.

\begin{figure}[H]
\begin{center}
\includegraphics[width=4in]{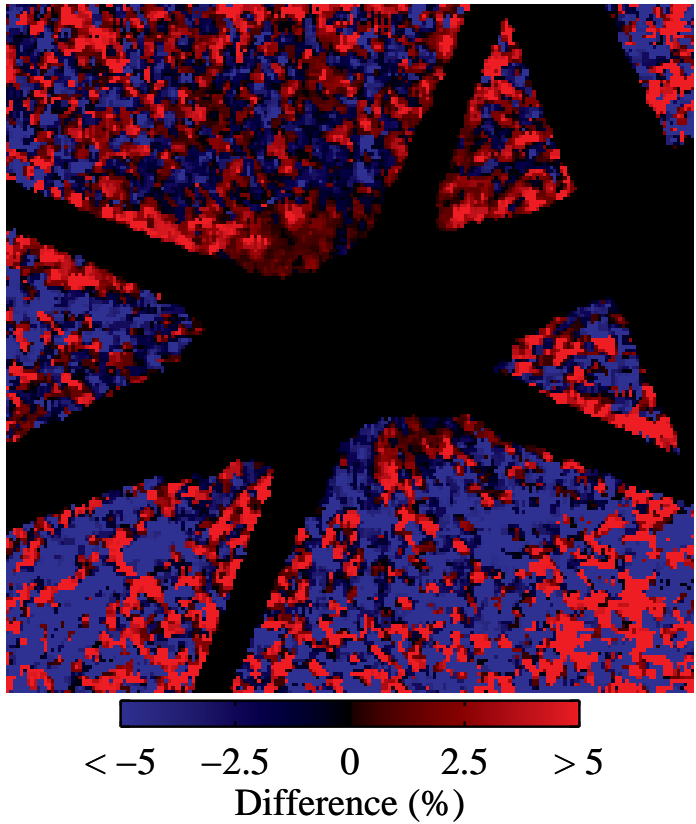}
\caption{Relative difference between a representative 11-image median and the 12-image median (central 200$\times$200 pixels).  PSF residuals in a single image can bias the flux of the 12-image median by $\sim5\%$ over scales $\sim10$ pixels or larger.    \label{residual_illustration_fig}}
\end{center}
\end{figure}

Our multi-roll observation technique reduces both the impact of temporal instabilities in the PSF structures (``breathing") and static PSF residuals that co-rotate with the instrument/telescope optics.  To estimate the remaining impact of PSF artifacts, we produced an 11-image median, subtracted it from the 12-image median, and divided by the 12-image median to produce a fractional residual map.  If the 12-image median is robust to these PSF residuals, then leaving out any one image should not greatly impact the final image.  Figure \ref{residual_illustration_fig} shows one such fractional residual map.  The fractional residuals, smoothed using a 3$\times$3 pixel median boxcar and displayed on a saturated scale for illustrative purposes, show correlated biases in the median by as much as $\sim5\%$ over scales $\sim10$ pixels.  Additionally, Figure \ref{residual_illustration_fig} shows that leaving out this particular image affects the NE-SW asymmetry at this level by enhancing the NE flux and reducing the SW flux.

\begin{figure}[H]
\begin{center}
\includegraphics[width=4in]{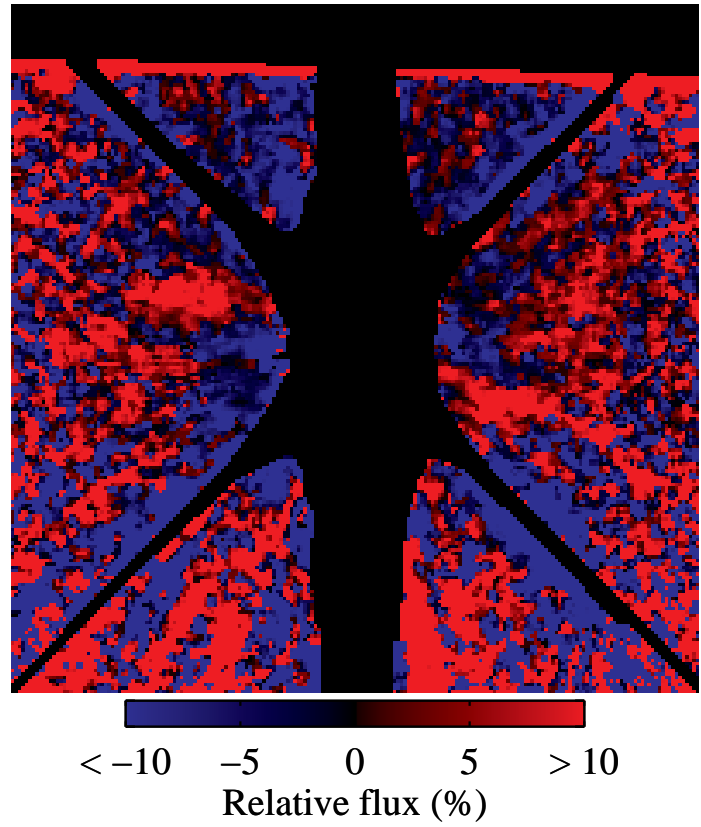}
\caption{Static PSF residuals common to all images in SIAF frame, smoothed by a 3$\times$3 boxcar median (central 200$\times$200 pixels).  We remove these residuals from each individual image to create an improved multi-roll median image.   \label{residual_fig}}
\end{center}
\end{figure}

\begin{figure}[H]
\begin{center}
\includegraphics[width=6in]{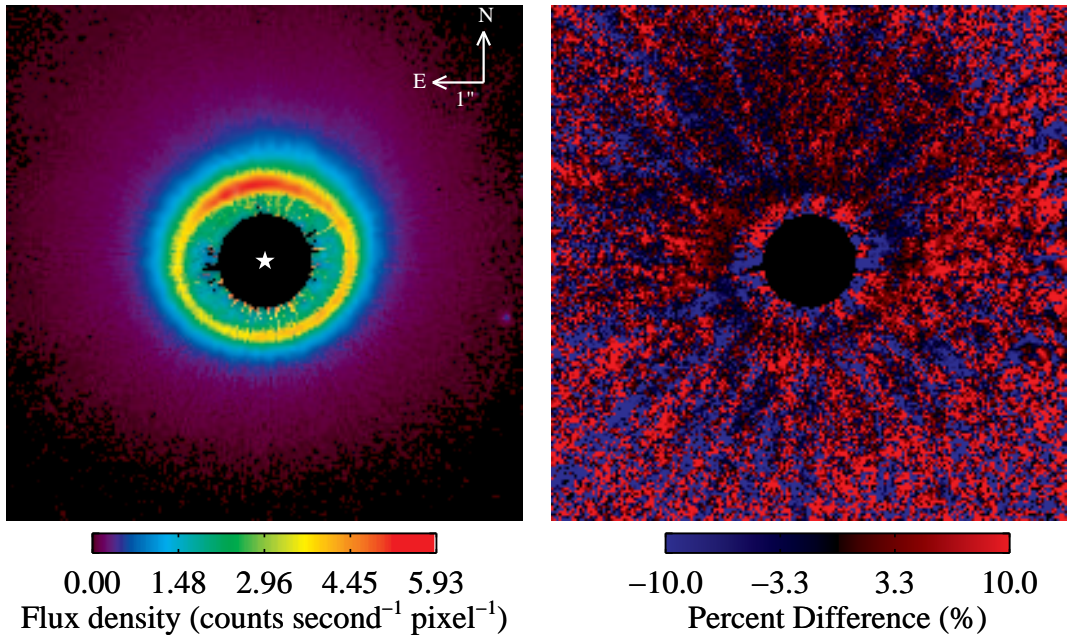}
\caption{\emph{Left:} Final multi-roll image of HD 181327 after processing it with one iteration of MRRR. \emph{Right:} Fractional updates to the image produced by MRRR.  \label{final_image_fig}}
\end{center}
\end{figure}

Although we cannot remove time-dependent PSF residuals with this method, we can better mitigate static PSF residuals.  We did this using the following ``multi-roll residual removal routine" (MRRR, pronounced ``myrrh"):
\begin{enumerate}
\item Form the 12-image median, oriented north-up.
\item Subtract the 12-image median from each individual masked image, oriented north-up, to form 12 disk-less residual images.
\item De-rotate each disk-less residual image to the SIAF-oriented frame.
\item Take the median of the 12 disk-less residual images to produce a map of static residuals.
\item Smooth the static residual map using a 3$\times$3 median boxcar to remove pixel-to-pixel noise, retaining only larger scale correlated residuals.
\item Subtract the smoothed, correlated residual map from each of the 12 images in the SIAF frame, rotate each to the north-up frame, and form a new residual-removed 12 image median.
\end{enumerate}
Figure \ref{residual_fig} shows the smoothed, correlated residual map for the inner 200$\times$200 pixels of our observations of the \hd\ disk (displayed on a saturated scale for illustration).  Figure \ref{final_image_fig} shows the final residual-removed 12 image median, and the fractional difference between the original 12-image median and the residual-removed 12 image median.  In this case, MRRR dims the ansae by $\sim5\%$ and brightens the SW side of the ring by $\sim5\%$.  Not surprisingly, the regions of the disk impacted most significantly are the regions with small values of $n_{\rm pix}$ (see Figure \ref{twelve_image_median_fig}).  We use the MRRR-corrected image for all subsequent analysis.

One could in principle run MRRR several times over, iteratively removing the correlated static residuals.  A preliminary implementation of an iterative MRRR appears to converge in only a few iterations, with higher order iterations predominantly brightening the SW side of the disk by an additional $\sim2\%$.  However, MRRR requires the correlated PSF residual amplitudes to be larger than the Poisson noise remaining in the disk-less images.  It is unclear to us at what point this assumption breaks down, so for this work we conservatively perform only a single iteration.

Modeling our MRRR-corrected image of the HD 181327 debris disk requires an understanding of the uncertainty in the PSF-subtracted flux in each pixel.  On average, the standard deviation of the flux measurements within each pixel greatly exceeds what is expected for Poisson noise alone, so the observations are dominated by systematic noise, as is typical for HST PSF template-subtracted imaging.  In principle, one could use the standard deviation in each pixel as an estimate of each pixel's uncertainty.  However, with a median $n_{\rm pix}= 6$ near the ring's peak flux, a large number of pixels will, by chance, have a standard deviation that is much smaller than the true systematic uncertainty.  These pixels will dominate the error-budget of any model fitting routine.  One could arbitrarily mask off pixels whose S/N exceeds some threshold, but that would remove useful data from the fit.

As an alternative, we assume the dominant systematic noise is only a function of stellocentric angular distance.  Because such noise is not dependent on disk brightness, we separately subtract each of our original 12 images from the MRRR-corrected image, then calculate the standard deviation of this set of differences as a function of stellocentric distance--this is equivalent to the standard deviation of the unsmoothed residuals.  We then divide the azimuthally-symmetric standard deviation by the greater of $\sqrt{n_{\rm pix}-1}$ and unity to account for the number of observations per pixel, $n_{\rm pix}$, while avoiding a systematic underestimate of the uncertainty given small values of $n_{\rm pix}$.  Note that we base our uncertainty estimates on the first iteration residuals, which contain the PSF residuals removed by MRRR; our conservative uncertainty estimate budgets for the changes made by MRRR.

\section{Results and Analysis}
\label{results}

The MRRR-corrected image of \hd, shown in Figure \ref{final_image_fig}, exhibits several asymmetries that are immediately noticeable.  The NE side of the disk exhibits a peak in the surface brightness, approximately $30\%$ brighter than the SW side of the disk.  There is also a NW-SE asymmetry, with the NW side of the disk approximately $10\%$ brighter than the SE side.  Additionally, the disk appears more radially extended toward the N than toward the S.

It is not immediately obvious whether these asymmetries are due to local dust density enhancements or a combination of geometric and scattering effects.  A belt with significant density enhancements may suggest the presence of a nearby perturbing planet---or that a massive collision recently occurred.  Simulations of planets perturbing debris disks have shown that planets are capable of producing belts with sharp inner edges \citep[e.g.][]{2006MNRAS.372L..14Q,2009ApJ...693..734C, 2014ApJ...780...65R} as well as azimuthal asymmetries in the dust density \citep[e.g.][]{1999ApJ...527..918W,2003ApJ...588.1110K,2012A&A...544A..61E}.  

The scattered-light asymmetries we observe in the \hd\ disk are primarily located along the minor and major axes of the projected belt, suggesting that they may be due, at least in part, to geometric and scattering effects.  Here we seek to determine whether geometric and scattering effects \emph{alone} can explain these apparent morphological asymmetries, or if an actual dust density enhancement is necessary.  To do so, we assume that the disk is infinitely thin and flat (an assumption we address in Section \ref{scale_height_section}), deproject the disk to a ``face-on" viewing geometry, and remove any asymmetries that can be explained by geometric and scattering effects.  Any significant asymmetries that remain must therefore be density enhancements or deficits, or are signs that the disk is not flat.

To deproject the observations and examine the face-on optical depth of the disk we employed the following process:
\begin{enumerate}
\item Fit the observed (projected) \hd\ dust belt with an ellipse
\item Deproject the ellipse to obtain the true orbital ellipse and disk geometry
\item Using the true orbital ellipse values in the projected image plane, correct for the $1/r^2$ illumination factor
\item Determine the best fit scattering phase function and divide it out of the image
\item Deproject the image to produce a final face-on optical depth image
\item Examine the optical depth for remaining density asymmetries
\end{enumerate}
Below we discuss this process step by step.  In practice, steps 1 -- 3 were incorporated into their own iterative subroutine which we describe below.  We will repeatedly refer to Figure \ref{deprojection_figure}, which illustrates each step of this process.

\subsection{Projected ellipse fitting \label{ellipse_fit_section}}

There are many ways to fit an ellipse to the \hd\ dust ring shown in Figure \ref{final_image_fig}.  First, we must choose a metric that defines the ellipse, e.g. the location of the belt's peak flux, an isophote, the belt's inner edge, etc.  Ideally we'd choose a metric that reflects the underlying density distribution, e.g. the peak density of the belt.  However, the radial location of the peak surface brightness does not correspond to the radial location of the peak density, because the differential $1/r^2$ illumination factor shifts the apparent peak location (an especially important factor for eccentric debris rings).  We must correct for the $1/r^2$ illumination factor to fit the orientation of the disk, but we must know the orientation of the disk to correct for the $1/r^2$ illumination factor.

In light of this catch-22, we developed a method to fit the illumination-corrected image by iterating steps 1 -- 3 of our disk deprojection procedure.  First, we guessed the orientation of the disk (e.g. circular and face-on).  We then generated a map of $1/r^2$ and used it to remove the stellar illumination.  We then recorded the coordinates of the illumination-corrected projected belt maximum, fit these coordinates with an ellipse, and deprojected the ellipse to obtain a new disk orientation.  We iterated this process, using updated $1/r^2$ maps each time.  It's unclear whether this procedure should converge in all cases, but in the case of \hd, we found that this process converged in $\sim3$ iterations regardless of our initial guess of the disk orientation.

We used polar coordinates to measure the projected radius of the peak illumination-corrected surface brightness, which we refer to loosely as the surface density.  We calculated on-sky $\rho$ and $\phi$ values for each pixel's center relative to the star location.   We divided the surface density image into 172 wedges with angular size of 2.1$^{\circ}$ centered on the star, chosen such that the arc length of a wedge is $\sim1$ pixel at the location of the projected belt's semi-minor axis.  

For each wedge, we measured the sub-pixel radial location of the ring maximum by fitting a Gaussian to the radial profile near the peak surface density.  To find the ideal number of pixels to fit, we made separate Gaussian fits to the nearest 7-17 pixels and chose the fit with the most certain peak location.  This method produces a set of $\rho_{\rm peak}$, the best fit radial surface density peak, at each angular location, $\phi_{\rm peak}$.

We note that the uncertainty of the peak location of each wedge ultimately controls the uncertainty in the belt's semi-major axis, eccentricity, and orientation.  Thus, it is important to estimate the uncertainty in the peak robustly.  To do this, we used a Monte Carlo procedure.  Each time we fit the data with a Gaussian, we performed 100 Monte Carlo trial fits.  For each of these 100 Monte Carlo trials, we added a random flux to each pixel, drawn from a normal distribution with standard deviation equal to the pixel's flux uncertainty, and refit with a Gaussian.  The uncertainty in the peak location was then set equal to the standard deviation of the Gaussian peak locations from the 100 Monte Carlo trials.  

We then ran through a fine grid of ellipse parameters to search for the best fit to the $(\rho_{\rm peak},\phi_{\rm peak})$ coordinates.  For each set of ellipse parameters we calculated a $\chi^2$ value by comparing each $(\rho_{\rm peak},\phi_{\rm peak})$ point with the nearest point to the ellipse, i.e. we minimized the perpendicular distance to each point weighted by the uncertainty.  Table \ref{ellipse_fits_table} lists the parameters of the best fit ellipse to the illumination-corrected belt maximum.  Here the primed quantities refer to the on-sky, projected ellipse.  We fit the semi-major axis, $a'$, eccentricity, $e'$, position angle, PA$'$, and the location of the ellipse center $(\Delta x', \Delta y')$ relative to the star.  To determine the 1$\sigma$ uncertainties, we normalized the $\chi^2$ values such that the minimum $\chi^2$ was equal to unity, then took the projection of the parameters for all models with normalized $\chi^2 < 2$.   We consider all projected ellipse fits with normalized $\chi^2 < 2$ to be acceptable fits, and analyzed all of these fits in the following sections.

\begin{deluxetable}{cccccc}
\tablewidth{0pt}
\footnotesize
\tablecaption{Best fit ellipses to the projected \hd\ debris belt \label{ellipse_fits_table}}
\tablehead{
\colhead{Method} & \colhead{$a'$ (pixels)/(AU)} & \colhead{$e'$} & \colhead{PA$'$ ($^{\circ}$)} & \colhead{$\Delta x'$ (pixels) / (AU)} & \colhead{$\Delta y'$ (pixels) / (AU)}
}
\startdata
\vspace{0.1in}
Maximum & $34.4^{+0.4}_{-0.4}$ / $90.5^{+1.1}_{-1.1}$  & $0.48^{+0.03}_{-0.03}$ & $101.2^{+4.6}_{-4.6}$ & $-0.34^{+0.35}_{-0.38}$ / $-0.89^{+0.92}_{-1.00}$ & $0.52^{+0.30}_{-0.30}$ / $1.37^{+0.79}_{-0.79}$\\
\vspace{0.1in}
Inner edge & $31.3^{+0.4}_{-0.4}$ / $82.3^{+1.1}_{-1.1}$ & $0.50^{+0.03}_{-0.03}$ & $98.5^{+3.7}_{-4.0}$ & $-0.28^{+0.34}_{-0.32}$ / $-0.74^{+0.89}_{-0.84}$ & $-0.11^{+0.26}_{-0.24}$ / $-0.29^{+0.68}_{-0.63}$\\
\enddata
\end{deluxetable}

To check whether our results depended strongly on the ellipse metric we chose, we repeated this procedure by fitting the inner edge of the belt.  To fit the inner edge, we first calculated the radial derivative of the illumination-corrected \hd\ image in the true disk plane, for which the maximum marks the belt's inner edge.  We then fit the maximum of the radial derivative using the techniques described above.  Table \ref{ellipse_fits_table} lists the results of fits to the inner edge of the belt.  With exception to the values of $a'$, which should not agree, and $\Delta y'$, the fits agree to within 1$\sigma$, suggesting that the disk is thin and flat near the brightest part of the observed ring.  The small discrepancy in the $\Delta y'$ values may be a result of the blurring of radial features along the disk minor axis due to a small non-zero disk scale height, which we address in Section \ref{scale_height_section}.

\subsection{Ellipse deprojection}

To deproject the ellipse fits, i.e. obtain the true geometry and orientation of the disk, we used the Kowalsky method, as described by \citet{1930MNRAS..90..534S}.  This analytic method transforms the parameters describing an apparent ellipse with a center offset from the star ($a'$, $e'$, PA$'$, $\Delta x'$, $\Delta y'$) into a set of unique, deprojected ellipse parameters ($a$, $e$, $\omega$, $i$, $\Omega$) describing both the geometry and orientation of the true ellipse.  Here $a$ is the true semi-major axis, $e$ is the true eccentricity, $\omega$ is the argument of pericenter, which defines the axis of inclination in the plane of the disk, $i$ is the inclination, and $\Omega$ is the longitude of the ascending node, which defines the angle of the axis of inclination on the sky.

We applied this method to all of the ellipse fits within 1 $\sigma$ of the best fit values listed in Table \ref{ellipse_fits_table}.  Table \ref{deprojected_fits_table} lists the resulting deprojected values.  We constrain the disk inclination to $28.5\pm2^{\circ}$, consistent with previous estimates \citep{2006ApJ...650..414S}, and marginally constrain the eccentricity to a non-zero value of $0.02\pm0.01$.  This small true eccentricity results in a poorly constrained $\omega$.  As a result, the pericenter of the \hd\ disk could be located anywhere in the SW quadrant of Figure \ref{final_image_fig}.

\begin{deluxetable}{cccccc}
\tablewidth{0pt}
\footnotesize
\tablecaption{Deprojected ellipse parameters for the \hd\ debris belt \label{deprojected_fits_table}}
\tablehead{
\colhead{Method} & \colhead{$a$ (AU)} & \colhead{$e$} & \colhead{$\omega$ ($^{\circ}$)} & \colhead{$i$ ($^{\circ}$)} & \colhead{$\Omega$ ($^{\circ}$)}
}
\startdata
\vspace{0.1in}
Maximum & $90.5^{+1.1}_{-1.1}$  & $0.02^{+0.01}_{-0.01}$ & $-70^{+32}_{-33}$ & $28.5^{+2.1}_{-2.0}$ & $11.2^{+4.6}_{-4.6}$\\
\vspace{0.1in}
Inner edge & $82.3^{+1.1}_{-1.1}$ & $0.01^{+0.01}_{-0.01}$ & $16^{+164}_{-196}$ & $30.3^{+1.9}_{-2.0}$ & $8.5^{+3.7}_{-4.0}$\\
\enddata
\end{deluxetable}

\begin{figure}[H]
\begin{center}
\includegraphics[width=4in]{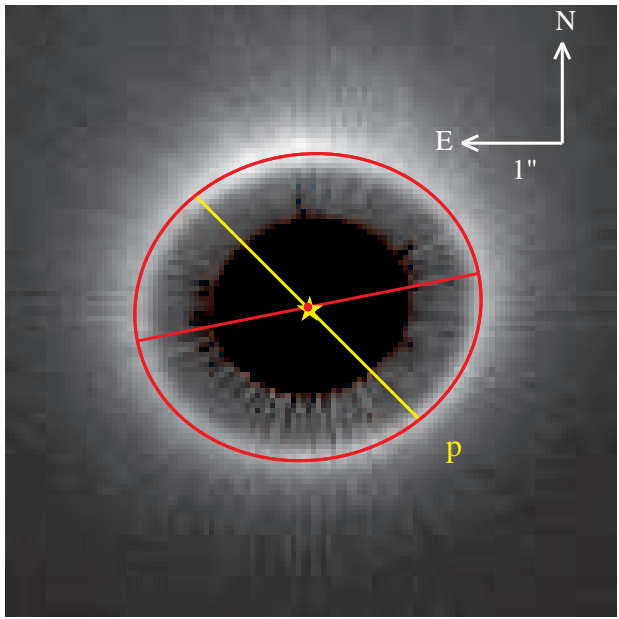}
\caption{Best fit ellipse to the belt maximum.  The projected major axis and true major axis are shown in red and yellow, respectively, with periastron at point ``p."  The stellar location is marked with a yellow star.  \label{geometry_figure}}
\end{center}
\end{figure}

The red ellipse in Figure \ref{geometry_figure} shows the best fit to the \hd\ illumination-corrected belt maximum.  The red line illustrates the major axis of the projected ellipse and the yellow line marks the true major axis with periastron marked with a ``p."  The line of nodes (the axis of inclination of the disk) is nearly coincident with the projected major axis, as expected for a nearly-circular ring; the line connecting forward to back scattering is approximately perpendicular to the projected major axis.  The center of the ellipse and the star are marked with a red dot and a yellow star, respectively.

Table \ref{deprojected_fits_table} also lists the deprojected ellipse parameters for fits to the inner edge of the belt.  In the case of an infinitely thin and uniform disk, we expect ellipses fit to the belt maximum and the belt inner edge to agree within their mutual uncertainties, except for $a$.  A discrepancy would suggest that the radial distribution of dust varies significantly, that the disk is not flat, or that the disk has an opening angle of more than a few degrees. As shown in Table \ref{deprojected_fits_table}, fits to the illumination-corrected belt maximum and the illumination-corrected inner edge give consistent values for all parameters and have similar uncertainties; a flat, thin disk appears to be a reasonable approximation for the \hd\ debris belt.

\subsection{Illumination correction}

We next corrected for the $1/r^2$ stellar illumination factor.  We distributed $10^7$ particles uniformly in azimuth and logarithmically in $r$ from 50 to 800 AU.  Using the deprojected ellipse parameters from the first line of Table \ref{deprojected_fits_table}, we then rotated the 3-D positions by $\omega$, $i$, and $\Omega$, and binned the particles into pixels based on their on-sky ($x$,$y$) coordinates, the pixel scale of $50.77$ mas, and a distance to \hd\, of $51.8$ pc.  Finally, we calculated the average $1/r^2$ value in each pixel to produce a $\langle 1/r^2\rangle$ map, which we divided into the \hd\ STIS image.
 
\begin{figure}[H]
\begin{center}
\includegraphics[width=6.5in]{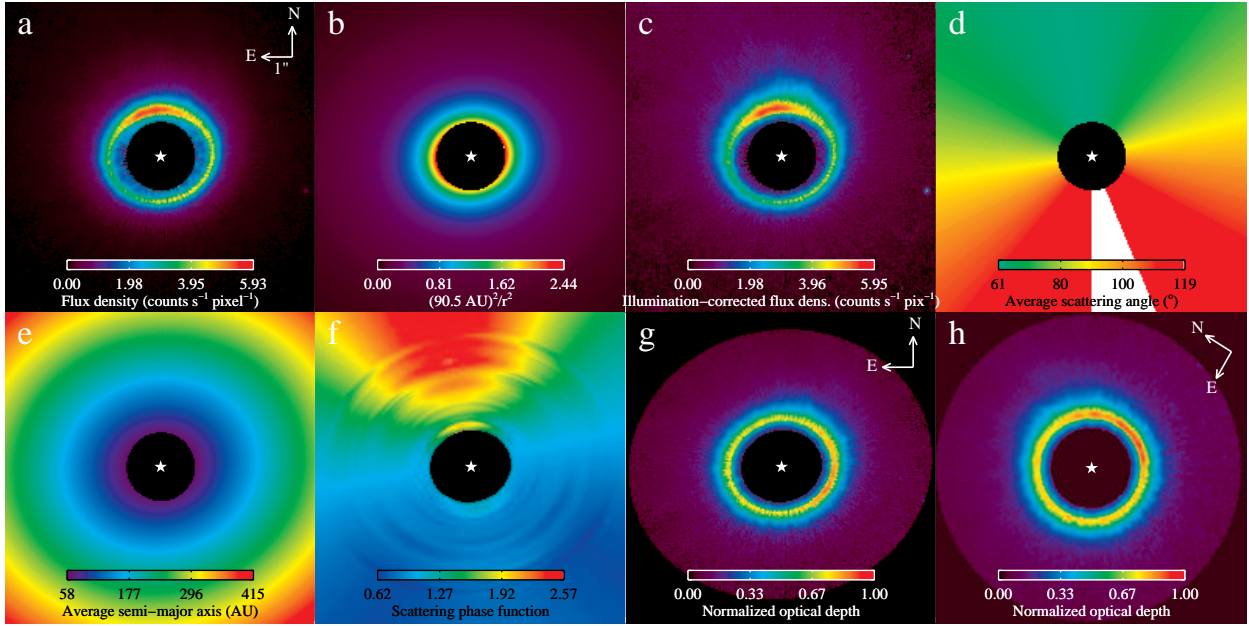}
\caption{Deprojection procedure. a) Original STIS image. b) $\langle 1/r^2\rangle$ map normalized to the birth ring distance of $90.5$ AU.  c) (a) divided by (b).  d) Map of the average scattering angle.  e) Map of the average semi-major axis.  f) Best fit scattering phase function map.  g) (c) divided by (f). g) (h) Deprojected optical depth with the belt's pericenter located directly to the right of the star.  The stellar location is marked in each image with a white star.  \label{deprojection_figure}}
\end{center}
\end{figure}

Figure \ref{deprojection_figure} shows this portion of the deprojection process.  Panel (a) shows the initial \hd\ STIS image.  Panel (b) shows the $\langle 1/r^2\rangle$ map for the deprojected ellipse fit listed in the first line of Table \ref{deprojected_fits_table}, normalized to the radial peak at $90.5$ AU.  Panel (c) shows the illumination-corrected image, given by the \hd\ STIS image divided by the $\langle 1/r^2\rangle$ map.

\subsection{The scattering phase function(s) and density distribution\label{spf_section}}

The scattering phase functions (SPFs) of disks are commonly approximated using a Henyey-Greenstein (HG) SPF, given by
\begin{equation}
	p\left(\theta\right) = \frac{1}{4\pi}\frac{1-g^2}{[\,1+g^2-2g\cos{\theta}\,]^{3/2}},
\end{equation}
where $\theta$ is the scattering phase angle and $g$ is the HG asymmetry parameter, ranging from -1 for perfect backscattering to 1 for perfect forward scattering.  Small dust grains are known to be forward scattering, so typically $g>0$ for debris disks.  However, this function is typically used out of expedience, not fidelity.  SPFs predicted by Mie theory do not resemble HG functions in many cases, and HG SPF fits to observed debris disks produce $g$ values much less than is expected for micron sized grains \citep[e.g.][]{2005Natur.435.1067K,2006ApJ...650..414S,2008ApJ...673L.191D,2011ApJ...743L...6T}.  Additionally, the SPF of the zodiacal dust cloud is significantly flatter near a scattering phase angle of $90^{\circ}$ than predicted by a single forward scattering HG phase function \citep{1985A&A...146...67H}.

Instead of using an analytic phase function, we fit an empirical scattering phase function to the data.  At a given semi-major axis, $a$, the illumination-corrected flux, $F'$, shown in Figure \ref{deprojection_figure}c is proportional to
\begin{equation}
	F'\!\left(a,\theta\right) \propto \int \frac{\mathrm{d}N\!\left(a,\theta\right)}{\mathrm{d}s}\, s^2\,  Q_{\rm sca}\!\left(s\right)\, p\!\left(\theta,s\right) \mathrm{d}s,
\end{equation}
where $\mathrm{d}N\!\left(a,\theta\right)\!/\mathrm{d}s$ is the differential number of grains of size $s$, $Q_{\rm sca}\!\left(s\right)$ is the scattering efficiency, and $p\!\left(\theta,s\right)$ is the scattering phase function.  In the case of a uniform, unperturbed disk with small eccentricity, the size distribution is constant along a given semi-major axis, i.e. independent of $\theta$.  As a result, the scattering phase function averaged over grain size
\begin{equation}
	p\left(a,\theta\right) = \frac{ \int \frac{\mathrm{d}N\!\left(a\right)}{\mathrm{d}s}\, s^2\,  Q_{\rm sca}\!\left(s\right)\, p\!\left(\theta,s\right) \mathrm{d}s }{\int \frac{\mathrm{d}N\!\left(a\right)}{\mathrm{d}s}\, s^2\,  Q_{\rm sca}\!\left(s\right)\mathrm{d}s}
\end{equation}
is a function of $a$ and $\theta$.  Therefore we can determine an empirical SPF by fitting the illumination-corrected flux as a function of $a$ and $\theta$.

Following the same procedure described above for the $\langle 1/r^2\rangle$ map, we created projected maps of the average scattering phase angle $\langle \theta \rangle$ (Figure \ref{deprojection_figure}d) and the average semi-major axis $\langle a \rangle$ (Figure \ref{deprojection_figure}e).  For a given value of $a$, we then selected those pixels with $|\langle a \rangle - a| < 2.63$ AU (1 pixel width) and fit the illumination-corrected flux as a function of $\langle \theta \rangle$.

\begin{figure}[H]
\begin{center}
\includegraphics[width=4in]{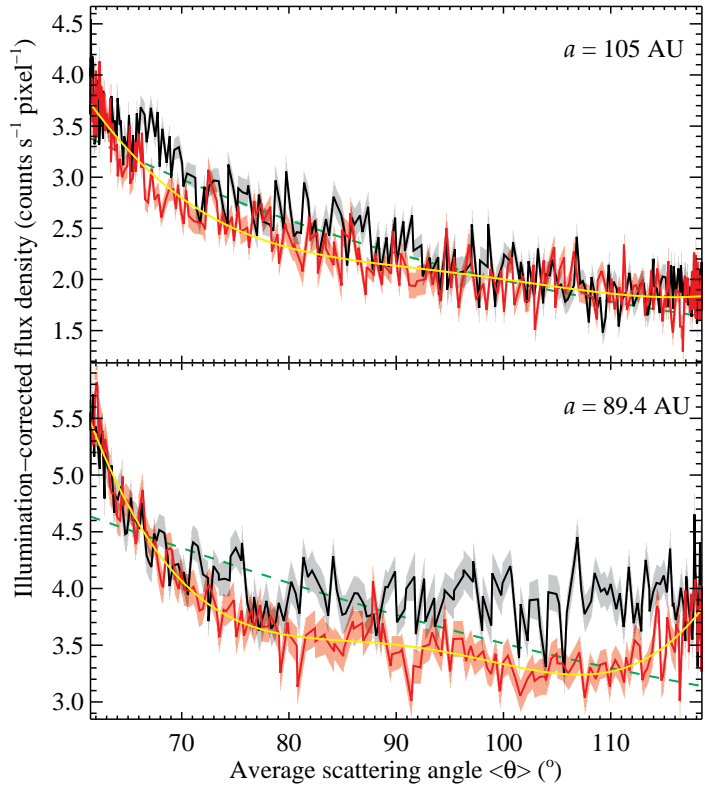}
\caption{Illumination-corrected flux as a function of scattering phase angle at and exterior to the belt maximum (lower and upper panels, respectively). The SE and NW halves of the disk are shown in red and black, respectively.  The yellow line shows the best fourth degree polynomial fit to the scattering phase function.  The dashed green line shows the best fit Henyey-Greenstein phase function, a poor fit to the observed variation.\label{spf_figure}}
\end{center}
\end{figure}

Figure \ref{spf_figure} shows the illumination-corrected flux as a function of $\langle \theta \rangle$ for two values of $a$.  Here we used the line joining the minimum and maximum scattering phase angles in Figure \ref{deprojection_figure} to divide the disk into a SE half (shown in red) and a NW half (shown in black).  In the case of a uniform disk, the SE and NW scattering phase functions should be identical and this is approximately true at $a=105$ AU, as shown in the top panel of Figure \ref{spf_figure}.  However, the bottom panel of Figure \ref{spf_figure} shows that near the belt maxium, $a=89.4$ AU, the two halves are asymmetric.  Under the assumption of a flat, thin disk, \emph{such asymmetries cannot be explained by additional projection or scattering effects and form the foundation by which we identify density asymmetries.}

We fit the flux measurements as a function of scattering phase angle with a fourth degree polynomial, shown in yellow in Figure \ref{spf_figure}, to obtain an empirical scattering phase function.  This function could be fit to the flux from the SE half of the disk, the NW half of the disk, or both halves simultaneously.  In the case of \hd\, the SE disk flux is consistently less than the NW flux.  We chose to interpret any detected asymmetries as density enhancements, so we fit the empirical scattering phase function to the flux from the SE half of the disk only.

\begin{figure}[H]
\begin{center}
\includegraphics[width=4in]{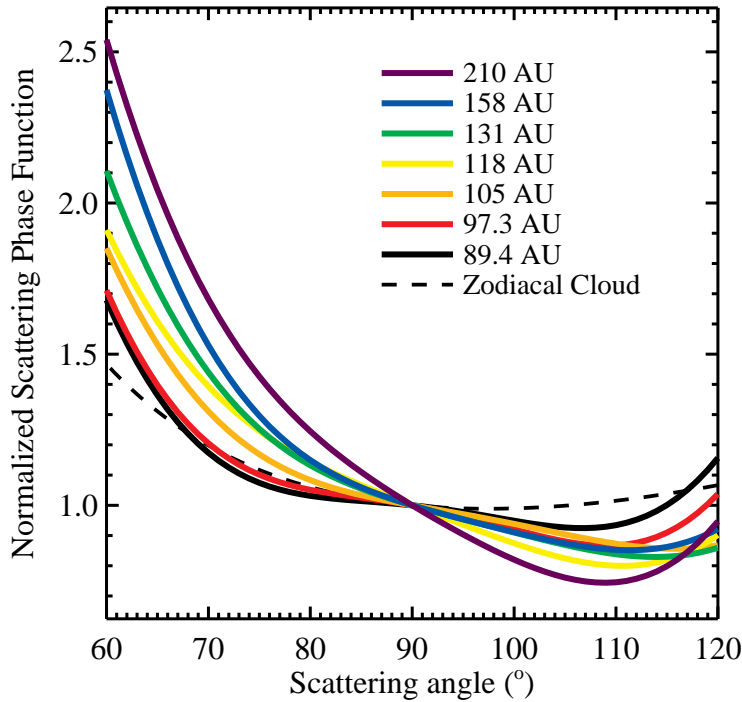}
\caption{The empirically derived scattering phase function as a function of $a$.  The functional behavior is consistent with Mie theory predictions for smaller grains at larger circumstellar distances. \label{spf_vs_a_figure}}
\end{center}
\end{figure}

Repeating this procedure for $a$ ranging from 71 AU to 263 AU in steps of $\Delta a = 2.63$ AU (1 pixel), we obtained the SPFs shown in Figure \ref{spf_vs_a_figure}.  Because grains beyond the parent body ring should be size-sorted (see Section \ref{spf_size_seg_section}), with $s$ decreasing with increasing $a$, the normalization of the SPF at a given $a$ is degenerate with the average $Q_{\rm sca}$ and surface density at a given $a$.  Additionally, the normalization of a given SPF strongly depends on the behavior of the SPF at small $\theta$, but our observations are limited to $60^{\circ} \lesssim \theta \lesssim 120^{\circ}$ given the disk inclination $\sim30^{\circ}$.  Thus, we chose to normalize each SPF at $\theta=90^{\circ}$, which maintains the radial profile along the $\theta=90^{\circ}$ line.  We note that, in the end, we present any detected density asymmetries in terms of their detection confidence, which is independent of the normalization.

Using the empirical scattering phase functions and the $\langle \theta \rangle$ maps shown in Figure \ref{deprojection_figure}d, we created the scattering phase function map shown in Figure \ref{deprojection_figure}f.  We divided the illumination-corrected surface brightness image by the scattering phase function map to produce the projected optical depth shown in Figure \ref{deprojection_figure}g.  Finally, rotating by $\Omega$ to align the longitude of nodes along the $x$ axis, stretching the image vertically by $1/\cos{i}$, and rotating by $\omega$ to place the periastron on the positive $x$ axis gives the face-on optical depth shown in Figure \ref{deprojection_figure}h.

\begin{figure}[H]
\begin{center}
\includegraphics[width=6.5in]{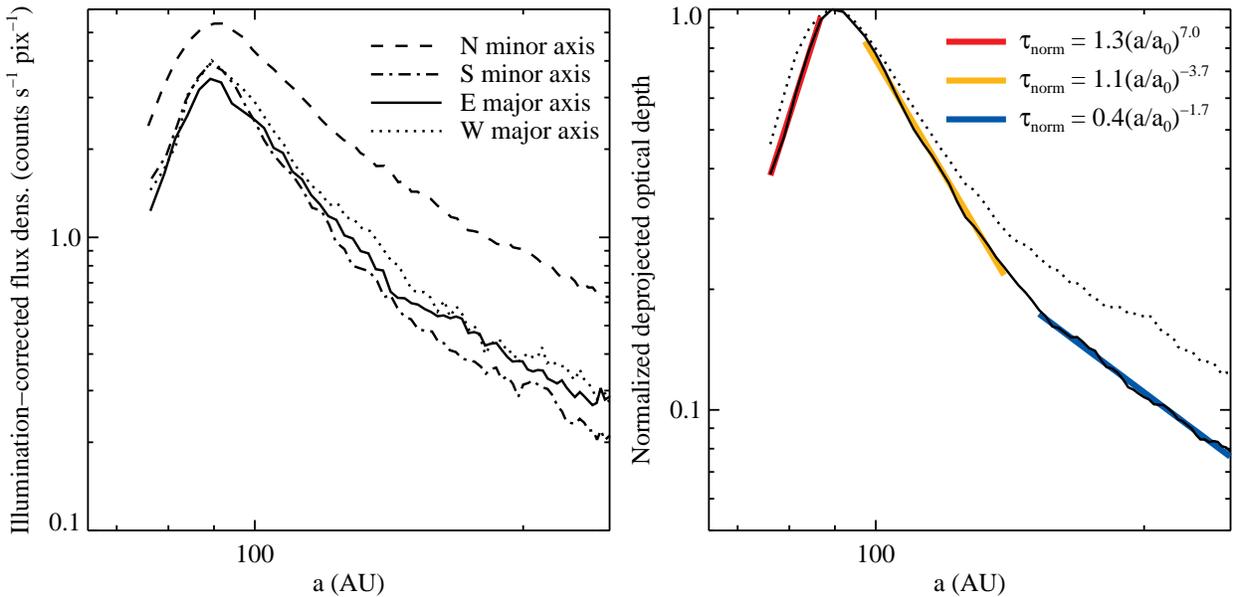}
\caption{\emph{Left:} Semi-major axis profile of the illumination-corrected flux density, median binned using a width $\Delta a=1$ pixel, along the apparent major and minor axes. \emph{Right:}  Median semi-major axis profile of the deprojected optical depth shown in panel (h) of Figure \ref{deprojection_figure} (solid black line) along with 3 power law fits (gray), where $a_0 = 90.5$ AU.  The dotted line shows the deprojected optical depth profile when normalizing the empirical SPF to a scattering angle of $60^{\circ}$; the FWHM of the radial optical depth profile is degenerate with a radial-varying SPF. \label{radial_profile_fig}}
\end{center}
\end{figure}

Figure \ref{radial_profile_fig} shows ``radial" profiles for the \hd\ disk, over the region for which we are confident PSF residuals are insignificant.  The left panel shows cuts along the major and minor axes of the illumination-corrected flux density as a function of semi-major axis using a median binning with width $\Delta a=1$ pixel.  The FWHM of the illumination-corrected flux varies depending on the cut, from $\approx25\%$ to $\approx40\%$.  Normalizing the SPF to a scattering angle of $90^{\circ}$ produces an optical depth profile with FWHM of $30\%$, while normalizing to a scattering angle of $60^{\circ}$ produces an optical depth profile with FWHM of $40\%$.  \emph{We do not know the correct normalization for the empirical SPF as a function of $a$, so the true radial dependence and FWHM of the optical depth is unknown.}  We attempted to normalize the SPFs by extrapolation to small scattering angles using a variety of appropriate functions; the resulting radial power law of the optical depth was wildly uncertain.  Thus, while the FWHM may help constrain the mass of a planet sculpting the inner edge of the disk in the case of a single SPF \citep{2014ApJ...780...65R}, if the SPF varies significantly with radial distance these constraints are less certain.

The right panel shows the median radial profile of the deprojected normalized optical depth when normalizing the SPF to a scattering angle of $90^{\circ}$.  The inner edge of the optical depth is $\propto a^{7.0}$.  The power law of the outer edge varies smoothly with semi-major axis.  From 95--140 AU, the optical depth $\propto a^{-3.7}$, while from 150--250 AU the optical depth is $\propto a^{-1.7}$.  The dotted line shows the median radial profile of the deprojected normalized optical depth when normalizing the SPF to a scattering angle of $60^{\circ}$.  The radial optical depth profile is degenerate with a radially-dependent SPF.

\subsection{Detected asymmetries}

The normalized optical depth in Figure \ref{deprojection_figure}h shows one obvious asymmetry: a density enhancement in the birth ring extending $\sim90^{\circ}$ counter-clockwise from periastron.  To reveal other less evident asymmetries, we took the difference between Figure \ref{deprojection_figure}h and its smooth counterpart, i.e. a uniform eccentric disk.  To create the deprojected optical depth for a uniform eccentric disk, we could simply calculate the median of the deprojected optical depth as  a function of $a$.  However, we chose to interpret any asymmetries as density enhancements, so we fit the optical depth as a function of $a$ using a smoothly varying polynomial, then set the optical depth of the smooth disk equal to the minimum of this function. 

\begin{figure}[H]
\begin{center}
\includegraphics[width=6.5in]{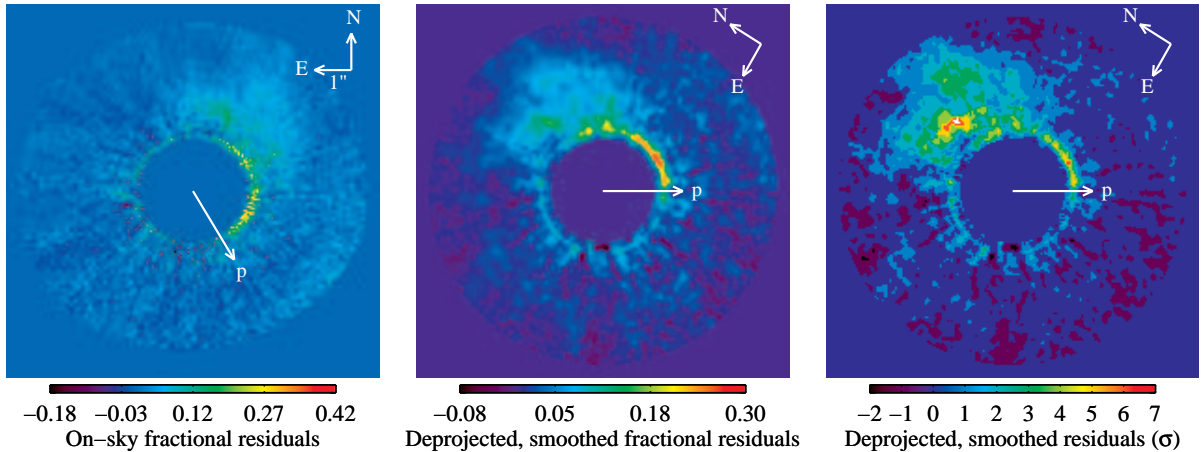}
\caption{Optical depth deviations from a uniform disk.  \emph{Left:} the on-sky projected plane. \emph{Middle:} smoothed residuals in the deprojected face-on plane.  \emph{Right:} detection confidence of the smoothed, deprojected residuals.  \label{asymmetry_figure}}
\end{center}
\end{figure}

Figure \ref{asymmetry_figure} shows the fractional optical depth residuals from what would be expected for a flat, thin, uniform disk.  The left panel shows these normalized residuals in the projected sky plane.  The middle panel shows the same residuals smoothed using a 3$\times$3 median boxcar in the deprojected, face-on plane with periastron located along the positive $x$ axis.  The right panel shows the detection confidence of the smoothed, deprojected residuals.

\section{Discussion}
\label{discussion}

We have detected asymmetric residuals in the \hd\, debris disk at $>3\sigma$ over many AU, suggesting the possibility of density enhancements/deficits.  However, to derive the residuals shown in Figure \ref{asymmetry_figure}, we assumed the disk was thin and flat.  Thus, there are three possible causes for the detected asymmetries.  First, the disk could indeed be thin and flat, implying that the detected asymmetries reflect true density enhancements or deficits.  Second, our assumption that the \hd\, disk is thin could be incorrect.  Third, our assumption that the \hd\, disk is flat could be incorrect.  We discuss each of these scenarios.

\subsection{Density enhancements in a thin, flat disk}

In the case of a thin, flat disk, \emph{the observed asymmetries cannot be explained by projection or scattering effects}; the detected asymmetries necessitate density asymmetries.  We note that we chose to interpret the asymmetries as density enhancements.  An equally valid, but less likely explanation is that the asymmetries represent deficits of material elsewhere in the disk.  Similarly, the detection confidence in the right panel of Figure \ref{asymmetry_figure} represents the confidence of \emph{an} asymmetry, not the confidence of whether an asymmetry is an excess as we have assumed, or deficit elsewhere in the disk.

\subsubsection{A massive collisional event}

Figure \ref{asymmetry_figure} shows that the density enhancement appears as an arc of material near periastron in the birth ring extending counter-clockwise and radially outward, with its angular size increasing with circumstellar distance.  The geometry of the detected asymmetry is consistent with models of small grains produced by high-mass collisional events \citep[e.g.][]{2007A&A...461..537G,2012MNRAS.425..657J,2013A&A...558A.121K,2014MNRAS.440.3757J}.  In these models, the smallest fragments with $\beta \gtrsim 0.5$ quickly leave the system on hyperbolic orbits.  However, the largest fragments, unaffected by radiation pressure, remain bound to the system on orbits that all cross at the location of the initial impact, creating a natural ``pinch" point.  These large grains routinely collide at this ``pinch," producing a continual outflow of small grains.  The lifetime of this asymmetry varies dramatically among models, from a few orbital periods to millions of years, and is largely dependent on the fraction of the dust mass associated with the presumed massive collision.

Interpreting the detected asymmetry as originating from a collisional event, we can place a lower limit on the mass of dust generated by the collision.  We multiplied the middle panel of Figure \ref{asymmetry_figure} by the optical depth peak value of $\tau_{\rm 0} = 2\times10^{-3}$, estimated from $L_{\rm IR}/L_{\star}$ for \hd\, \citep{2012A&A...539A..17L}, to obtain the absolute size-integrated optical depth.  We then multiplied by the pixel area, calculated from the observed image scale of $50.77$ mas pixel$^{-1}$, and an assumed distance to \hd\, of 51.8 pc \citep{2009A&A...501..941H} to obtain the absolute size-integrated product of the cross-section and scattering efficiency, $Q_{\rm sca}$.  We summed these values for all pixels with an S/N $> 2\sigma$ to calculate a total $Q_{\rm sca}$-weighted cross section associated with the collisional event.  We assumed the particles are well-described as porous mixtures of ice, amorphous silicate, and carbonaceous material, as determined from Mie theory by \citet{2012A&A...539A..17L}, and calculated the mass associated with the total $Q_{\rm sca}$-weighted cross section, where $Q_{\rm sca}$ was averaged over the STIS bandpass.  We assumed a size distribution $\mathrm{d}N/\mathrm{d}s \propto s^{-3.85}$, the expected size distribution of fragments from a recent disruptive collision \citep{2012ApJ...745...79L}, though the results are relatively insensitive to the size distribution assumed because $Q_{\rm sca}$ peaks near $s=1\;\mu$m. 

We find that the detected asymmetry requires $10^{20}$ kg of dust smaller than a few microns, equivalent to $1\%$ the mass of Pluto or $\sim10^{-4}$ the mass of the Kuiper Belt.   However, we only examined the asymmetric component of the optical depth and may have removed a significant fraction of the dust associated with this collisional event.  Therefore, the estimated dust mass represents a lower limit.

Calculating the total mass involved in the collision requires extrapolating the estimated mass over a narrow range of micron-sized dust grains to bodies kilometers in size.  The power law used for such an extrapolation is not well understood and may feature a number of breaks.  We therefore examine the lower limit on the total collisional mass, i.e. $100\%$ of the target mass is converted into grains smaller than a few microns.  Given the target mass of $10^{20}$ kg, the binding energy of the target can be approximated as the gravitational binding energy, or $10^{24}$ J.  The collisional energy is given by
\begin{equation}
E_{\rm col} = \frac{\mu}{2}v_i^2, 
\end{equation}
where $\mu = m_{\rm t}m_{\rm p}/(m_{\rm t} + m_{\rm p})$, $m_{\rm p}$ is the projectile mass, $m_{\rm t}$ is the target mass, and $v_i$ is the impact velocity.  Given the disk's small eccentricity, we assume the impact velocity is dominated by the scale height of the disk and set $v_i = (H/r) v_{\rm orbit}$, where $v_{\rm orbit} = 3.5$ km s$^{-1}$ is the orbital speed in the parent ring, and $H/r = 0.1$, the maximum value allowed by our analysis (see Section \ref{scale_height_section}).   Assuming half of the collisional energy is imparted to the target and $100\%$ of the imparted energy goes toward fragmenting the target, we estimate $m_{\rm p} \sim 10^{20}$ kg, i.e. the projectile and target mass are roughly the same mass.

Explaining the observed dust excess using a single collisional event becomes difficult when increasing the target mass beyond the lower limit of $10^{20}$ kg.  For example, assuming a size distribution $\mathrm{d}N/\mathrm{d}s \propto s^{-3.85}$ valid for all sizes, we estimate a total collisional mass of $10^{22}$ kg, roughly the mass of Pluto.  To catastrophically fragment a Pluto mass object, we must increase the impact velocity by 750 m s$^{-1}$.  We have ruled out values of $H/r>0.11$, so we must invoke more exotic scenarios, like a massive compact binary target or an unseen planet stirring the disk, to explain the catastrophic disruption of a Pluto mass object.

Alternatively, one could argue that the observed dust excess is not the result of a single recent collision, but a collisional avalanche initiated by a much less massive target.  \citet{2007A&A...461..537G} showed that the initial release of $10^{17}$ kg of dust in a disk with $\tau\sim10^{-3}$ can trigger a collisional avalanche with a total dust cross section that peaks at 200 times the initial cross section released.  Models of collisional avalanches  \citep{2007A&A...461..537G, 2013A&A...558A.121K} also predict morphologies qualitatively consistent with the observations.  A more detailed model of a collisional avalanche in the \hd\ disk is required to determine the probability of witnessing such an event.

\subsubsection{The SPF and size segregation}
\label{spf_size_seg_section}

An unperturbed narrow ring of parent bodies producing dust should naturally produce a size-sorted halo of dust exterior to the parent body ring.  Upon launch, a dust grain's orbit is modified by radiation and solar wind pressure.  The post-launch apastron distance of a dust grain increases with $\beta$, where $\beta$ is the ratio of radiation force to gravitational force on the dust grain.  For many materials and stellar systems, $\beta \propto s^{-1}$ is approximately valid all the way down to the blowout grain size.  Smaller grains achieve larger apastron distances and dominate the cross section density at larger circumstellar distances.

The empirical scattering phase functions in Figure \ref{spf_vs_a_figure} are consistent with this size sorting (orange through purple curves).  These curves show that over the range of observed scattering angles, the width of the forward scattering peak increases with semi-major axis.  This trend is consistent with what is expected from Mie theory if the dominant grain size decreases with increasing semi-major axis.

\begin{figure}[H]
\begin{center}
\includegraphics[width=4in]{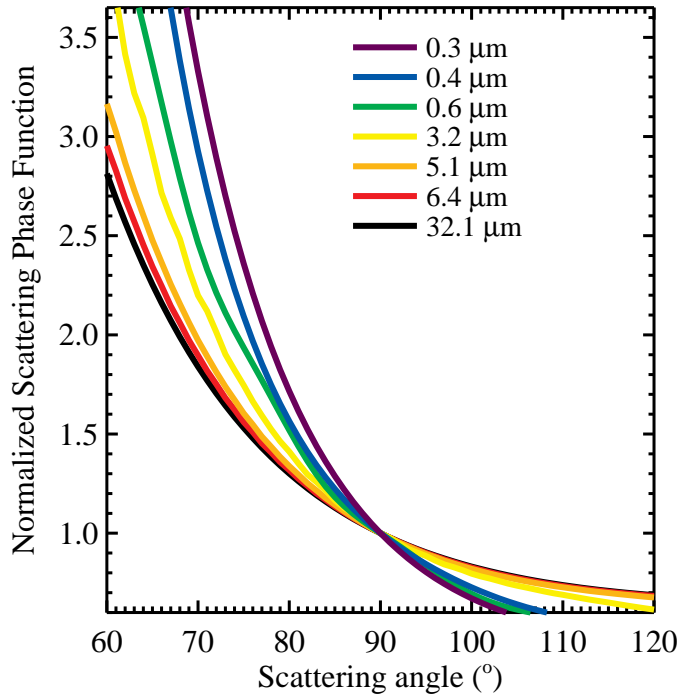}
\caption{SPF predicted by Mie theory for the best-fit composition of \citet{2012A&A...539A..17L}.  The trend suggests grain size decreases with increasing semi-major axis in the \hd\ debris disk, as predicted by dynamical debris disk models. \label{lebreton_spf_figure}}
\end{center}
\end{figure}

To illustrate how the empirical SPF suggests size segregation, Figure \ref{lebreton_spf_figure} shows the normalized scattering phase function predicted by Mie theory as a function of grain size for the porous mixture of ice, amorphous silicate, and carbonaceous material determined from fits to the spectral energy distribution (SED) of \hd\ as suggested by \citet{2012A&A...539A..17L}.  Although the magnitude of forward scattering for these grains does not agree with that shown in Figure \ref{spf_vs_a_figure}, the trend suggests size segregation in the \hd\ debris disk with particle size decreasing as semi-major axis increases.  This trend is a broad prediction of Mie theory; other compositions show similar trends.

While models show that larger grains should dominate the halo's optical depth closer to the birth ring, these same models predict that the trend stops at the outer edge of the birth ring.  As shown by \citet{2006ApJ...648..652S} and \citet{2014A&A...561A..16T}, the dominant grain size \emph{within} the birth ring should be approximately equal to the blowout size, i.e. the smallest grains in the system.  If this were the case for \hd, we would expect to see the trend in the empirical SPF reverse upon reaching the birth ring, i.e. the SPF at $a=89.4$ AU should be similar to that at $a=210$ AU.  Instead, the empirical scattering phase function in the parent body ring (black and red curves in Figure \ref{spf_vs_a_figure}) continues the observed trend, with an apparent decrease in the degree of forward scattering of the range of observed scattering angles.  Additionally, the degree of backscattering appears to increase and the SPF flattens near a scattering phase angle of $90^{\circ}$.

If the empirical SPF in the birth ring closely matches the true SPF, then there may be an absence of small grains in the birth ring.  Such a scenario is expected if a planetary companion orbits exterior to the birth ring.  As shown by \cite{2014A&A...561A..16T}, gravitational perturbations from an exterior planet can dynamically eject small grains, whose orbits are planet-crossing, before they can appreciably contribute to the optical depth at their periastron distance (the birth ring), while leaving the expected size sorting trend beyond the orbit of the planet unaffected.

The degree of forward scattering predicted by Mie theory, as shown in Figure \ref{lebreton_spf_figure}, does not closely match the empirical scattering phase function in the birth ring, even for very large grains.  Is the empirical scattering phase function measured in the birth ring reasonable for large debris disk grains?  The thin dashed line in Figure \ref{spf_vs_a_figure} shows the zodiacal cloud's derived scattering phase function (assuming $\nu=1$, see \citet{1985A&A...146...67H}), which is dominated by $\sim100\;\mu$m grains near 1 AU \citep{1985Icar...62..244G}.  Over the range of observed scattering angles, the zodiacal cloud's scattering phase function is less forward scattering than that observed for \hd; the scattering phase function from $100$ $\mu$m zodiacal cloud grains is also inconsistent with Mie theory, but consistent with the low degree of forward scattering observed in \hd.

\begin{figure}[H]
\begin{center}
\includegraphics[width=4in]{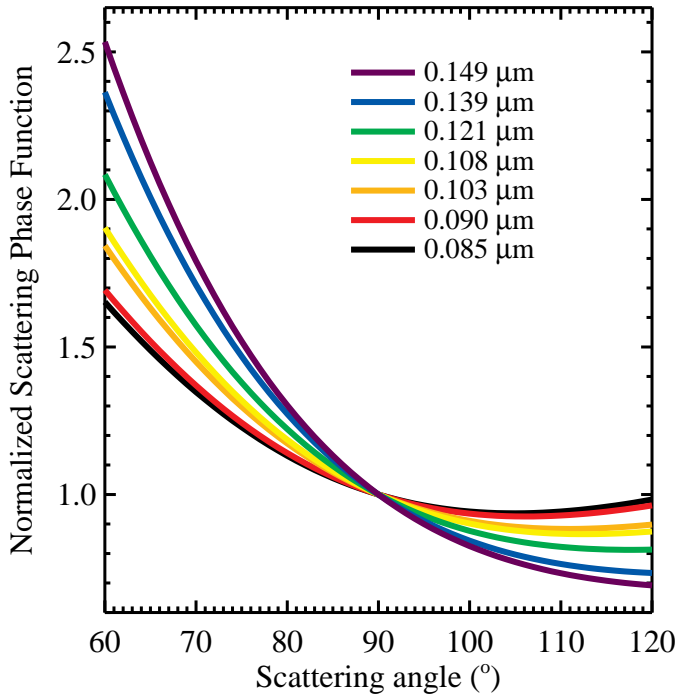}
\caption{Best-fit SPF predicted by Mie theory for the best-fit composition of \citet{2012A&A...539A..17L}.  These fits require sub-blowout-size grains to dominate the \hd\ optical depth. \label{lebreton_spf_small_grains}}
\end{center}
\end{figure}

Alternatively, the SPF predicted by Mie theory for $\sim0.1\; \mu$m grains fits the empirical SPF in the birth ring, and grains that increase in size with semi-major axis can fit the empirical SPF all the way out to 210 AU, as shown in Figure \ref{lebreton_spf_small_grains}.  However, such small grains are well below the blowout size for this system and do not survive long enough to dominate over the bound grains.  Further, for an unperturbed debris disk there is no known physical mechanism to create the trend of \emph{increasing} grain size with semi-major axis shown in Figure \ref{lebreton_spf_small_grains}.  Interestingly, the best fit SED model of \citet{2012A&A...539A..17L} also requires grains below the blowout size, $\sim0.8\; \mu$m.  There are three ways to interpret these results.  First, the best-fit porous composition of \citet{2012A&A...539A..17L} is correct, the optical properties of the \hd\ dust are truly dominated by grains below the blowout size, and we have a poor understanding of the dynamical behavior of such grains.  Second, the SPF predicted by Mie theory is not valid for the complex porous grains of \citet{2012A&A...539A..17L}.  Third, the SED modeling of \citet{2012A&A...539A..17L} needs to be revisited.  Given that \citet{2012A&A...539A..17L} assumed a single power-law size distribution valid at all points in the disk, we suspect the truth is a combination of all of these points.

All of the empirical scattering phase functions shown in Figure \ref{spf_vs_a_figure} deviate significantly from Henyey-Greenstein functions.  The dashed green lines in Figure \ref{spf_figure} show the best fit HG SPFs for $a=89.4$ and $a=105$ AU.  In each case, the slope of the empirical SPF is significantly larger than the best fit HG SPF at $\theta\sim60^{\circ}$.  Extrapolating this slope to even smaller, unobservable scattering angles could suggest that the \hd\ disk is far more forward scattering than previously thought.

\begin{figure}[H]
\begin{center}
\includegraphics[width=4in]{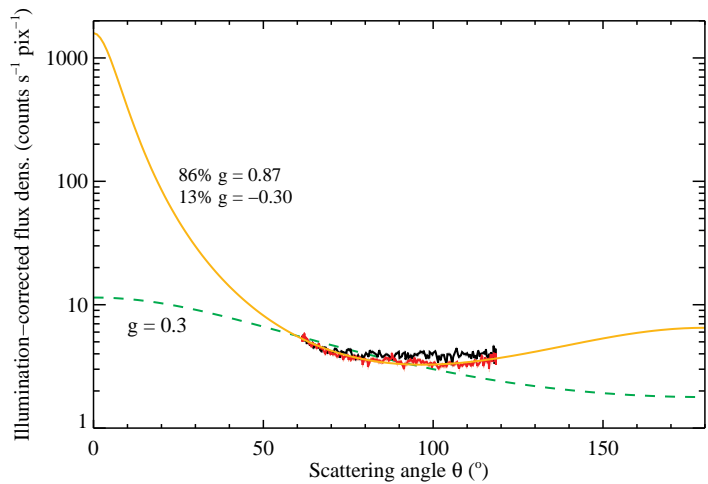}
\caption{Extrapolated fits to the SE illumination-corrected flux as a function of scattering phase angle at the location of the belt maximum.  The data is shown in red (SE) and black (NW).   The solid orange line shows the best fit 2-component HG SPF and the dashed green line shows best fit single HG with $g = 0.3$.  The apparently low albedo of the \hd\ disk may be entirely explained by a more strongly forward-scattering phase function (orange curve), though the extrapolation is not well-constrained.  \label{scatt_pf_extrap_fig}}
\end{center}
\end{figure}

The enhanced forward scattering that we find may help explain the apparently low albedo for the \hd\ disk.  Assuming a single HG scattering phase function with $g=0.3$, \citet{2012A&A...539A..17L} showed that the observed albedo for the \hd\ disk is lower than their model predictions by a factor of $\sim4$.  We extrapolated the empirical scattering phase function in the birth ring to all values of $\theta$ using a 2-component HG fit.  We found that two HG SPFs with $g_1=0.87$ and $g_2=-0.30$, weighted at $87\%$ and $13\%$, respectively, best fit the illumination-corrected SE flux (red curve in the lower panel of Figure \ref{spf_figure}), with a $\chi^2/\nu=2.2$.  The resulting SPF, shown in Figure \ref{scatt_pf_extrap_fig}, is significantly more forward scattering than a HG SPF with $g=0.3$.  As a consequence, the normalized empirical SPF is approximately 25\% that of a HG SPF with $g=0.3$ at $\theta=90^{\circ}$, eliminating the discrepancy noted in \citet{2012A&A...539A..17L}.  

Unfortunately the limited range of observable scattering phase angles prevents us from significantly constraining the empirical phase function at small scattering angles.  Considering all 2-component HG fits to the SE flux with $\chi^2/\nu$ within a factor of 2 of the best fit acceptable, values of $g_1 = [0.3,1.0]$; the extrapolated scattering phase function is statistically consistent with the albedo discrepancy noted in \citet{2012A&A...539A..17L}.  However, if the observed increase in flux at small scattering angles is real and due to the true scattering phase function, we regard fits with low $g_1$ with skepticism, as they do a poor job reproducing the SE flux at small scattering angles.

The discrepancy between the observed and modeled albedo from \citet{2012A&A...539A..17L} may also be an artifact of size segregation.   As previously noted, \citet{2012A&A...539A..17L} modeled the disk using a single grain size distribution at all circumstellar distances.  Instead, let's suppose the size distribution resembles that implied by the empirical SPF, with a paucity of small grains in the birth ring, and a halo dominated by small grains that decrease in size at larger circumstellar distances.  In this case the most efficient scatterers, the sub-micron grains, would not contribute as significantly to the scattered light image.  The thermal emission, however, will be dominated by the large grains in the parent body belt closer to the star.  As a result, the observed albedo will be lower than in the uniform size distribution case.

\subsubsection{Alternative density distributions}

We fit the illumination-corrected surface brightness of the SE half of the disk with an empirical SPF and discussed the implications of such an SPF above.  However, one could argue that the empirical SPF does not reflect the true SPF, and the SPF is degenerate with some non-uniform density distribution.  While strictly true, any observed surface brightness distribution can be attributed to a contrived density distribution.  For example, additional density enhancements near scattering angles of $90^{\circ}$ could cause the flatness of the empirical SPF in the birth ring.  In light of this, we limit ourselves to two relevant physical scenarios: 1) a birth ring that is in fact occupied by small dust grains but has a density enhancement that masks the true SPF, and 2) a debris disk whose true SPF everywhere is approximately equal to the empirical SPF in the birth ring.

To investigate the nature of the necessary density enhancements we deprojected the disk again, but altered the SPF in Panel F of Figure \ref{deprojection_figure}.  For the first case, in which we assume the birth ring is occupied by small dust grains, we assigned the empirical SPF from pixels with $208 < a < 213$ AU to all pixels with $86.8 < a < 100$.  For the second case, we assigned the empirical SPF from pixels with $86.8 < a < 92.0$ AU to all other pixels.

\begin{figure}[H]
\begin{center}
\includegraphics[width=5in]{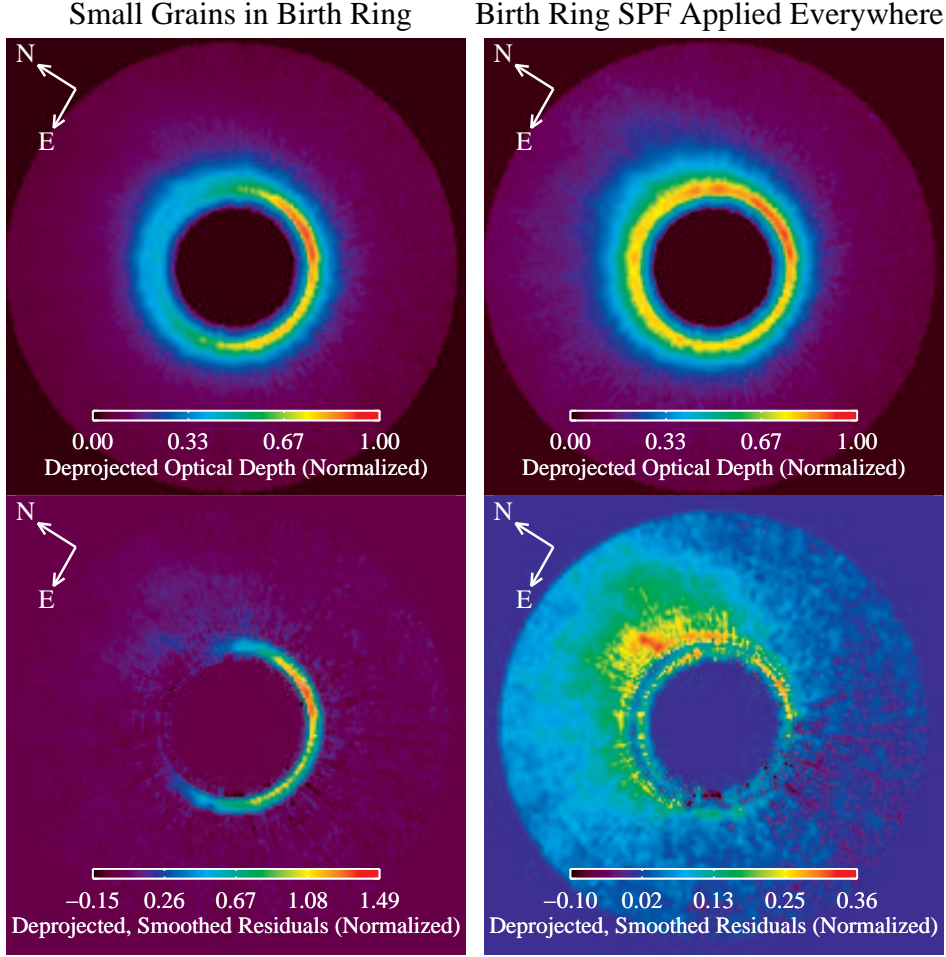}
\caption{Alternative deprojected optical depth distributions.  \emph{Left:} The required optical depth distribution if small grains dominate both the birth ring and outer disk, as expected from unperturbed disk theory (obtained by setting the SPF of the birth ring equal to the SPF of the outer disk). \emph{Right:} The required optical depth distribution if the empirical SPF in the birth ring is true everywhere. Both scenarios feature a greater degree of asymmetry than our preferred deprojection shown in Figure \ref{asymmetry_figure}. \label{alternative_densities_figure}}
\end{center}
\end{figure}

Figure \ref{alternative_densities_figure} shows the results of both of these scenarios in the deprojected frame with periastron pointing to the right.  For the true SPF in the birth ring to match the empirical SPF in the outer disk, as expected for an unperturbed narrow birth ring, there must be an additional density enhancement in the birth ring.  The left panels show that this density enhancement is $\sim100\%$ near periastron and is superimposed on top of the asymmetry previously detected.  The additional density enhancement must be symmetric and aligned, by chance, perpendicular to the line of nodes, a scenario we find unlikely.  

If we instead assign the empirical SPF of the birth ring to all other points in the disk, the required density enhancement in the outer disk increases significantly, as shown in the right panels of Figure \ref{alternative_densities_figure}.  Essentially the density distribution must make up for the lack of forward scattering in the birth ring's SPF.  This interpretation requires a slightly more massive collisional event, with $7.5\times10^{20}$ kg of dust less than a few microns in size, or the equivalent of $7.5\%$ the mass of Pluto.

\subsection{Constraints on the scale height of the \hd\, debris disk\label{scale_height_section}}

A non-zero scale height, which we have so far ignored, would affect the appearance of a debris disk in two ways.  First, for a disk with a vertical density distribution peaked at the midplane, the line of sight intersects more dust near the midplane when looking near the ansae; a non-zero scale height would brighten the disk along the ansae.  Second, a non-zero scale height would change the amount of flux received along the line joining forward to back-scattering.  Along this axis, the line of sight intercepts a wider range of circumstellar distances and scattering angles.

To constrain the scale height of \hd, we examined a sharp radial feature that a non-zero scale height would tend to blurr: the inner edge of the observed belt.  We fit a radial power law to the belt's flux as a function of projected semi-major axis at four locations within the inner edge ($71.0<a<86.8$ AU): the east and west disk ansae (along the axis of inclination), along the line of forward scattering, and along the line of back scattering.  The power law at the ansae should agree in the case of a uniform disk.  Unfortunately our observations are least certain in these regions, so we averaged the two power laws together.  We found radial power law exponents of $4.5\pm0.4$, $4.3\pm0.3$, and $4.0\pm0.5$ for the ansae, forward-scattering, and back-scattering sides of the disk, respectively.

We then produced Monte Carlo models of circularly symmetric disks including single-scattering radiative transfer for comparison.  We assumed a ``knife-edge" radial density profile, with $n(r) \propto r^{\beta}$ for $r < r_{\rm peak}$, and $n(r) \propto r^{\alpha}$ for $r > r_{\rm peak}$.  We investigated values of $\alpha$ in the range $[-12,-2]$ and $\beta$ in the range $[4,10]$, and $88.1 \le r_{\rm peak} \le 94.7$ AU.  We investigated disk scale heights $0 < H/r < 0.3$ and used a HG SPF to represent the SPF of the disk with $0 \le g \le 0.9$.  We produced a total of $1.6$ million models, each using 1 million particles, and convolved each model's image with a Tiny Tim point spread function, based on the stellar spectral template for an F6, B-V = +0.42 star.

We calculated the power laws at the inner edges of our models.  We found that only models with $H/r < 0.11$ produced profiles that were simultaneously within 1$\sigma$ of the measured values at the ansae, forward-scattering, and back-scattering sides of the disk.  Our constraint is consistent with the $H/r < 0.09$ constraint from\citet{2006ApJ...650..414S}.  

\begin{figure}[H]
\begin{center}
\includegraphics[width=4in]{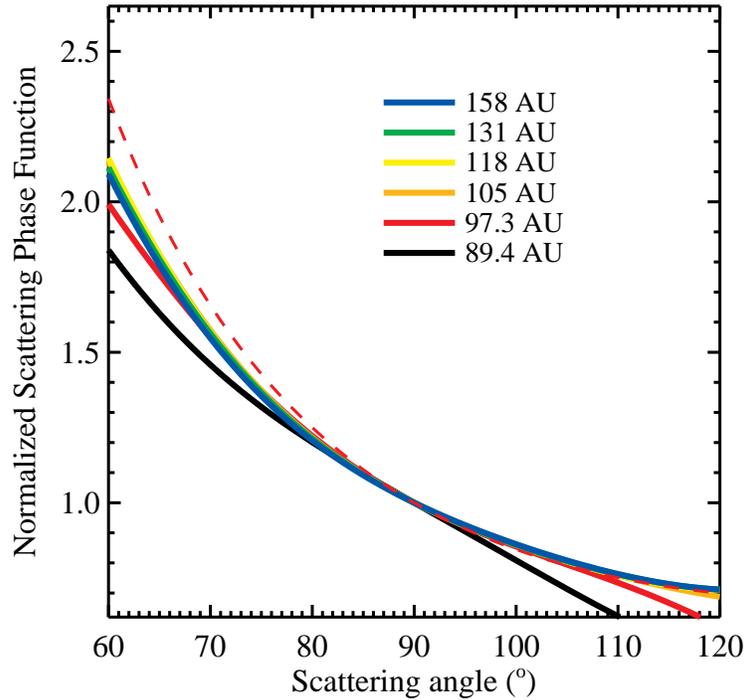}
\caption{SPF vs $a$ for our Monte Carlo model with a scale height of $H/r=0.3$.  A non-zero scale height decreases forward scattering in the birth ring as observed, but cannot produce the shape or spread of the empirical SPF. \label{model_spf_vs_a_figure}}
\end{center}
\end{figure}

Given our constraint on the scale height of the disk, could a non-zero scale height explain the variation in the empirical SPF observed for \hd?  To check, we calculated the apparent SPF for each of our Monte Carlo Models.  We found that only scale heights $\gtrsim0.3$ produced a significant decrease in the apparent forward scattering of the birth ring SPF, inconsistent with the constraints on the scale height.  Further, we were unable to qualitatively reproduce the curves shown in Figure \ref{spf_vs_a_figure} from our models, even for large scale heights.  Figure \ref{model_spf_vs_a_figure} shows an example of our model results for $H/r = 0.3$.  The models cannot produce SPFs with the observed spread in forward scattering from circumstellar distances of 105--210 AU.  Most critically, the models produced SPFs that were nearly tangential at $\theta=90^{\circ}$.  This is because a thick disk with a single radial power law beyond the birth ring reduces both forward and back scattering close to the disk, a trend that is not observed in \hd.  We conclude that a non-zero scale height alone cannot explain the variation in the empirical SPF.

The radial flux distribution in the outer disk is also inconsistent with the scenario of a thick disk with a single forward scattering phase function.  Power law fits to the flux, $f$, in the region $105 < a < 131$ AU reveal $f \propto a^{-5.6}$ along the ansae, but $f \propto a^{-5.1}$ in the direction of forward scattering and $f \propto a^{-6.9}$ in the direction of back scattering.  This trend is found regardless of the inner and outer boundaries of the fit as long as $a > 90.5$ AU, and is robust within the range of acceptable ellipse inclinations.  Our Monte Carlo models reveal that a thick disk with a single forward scattering phase function should have a radial power law that is \emph{steeper} in the direction of forward scattering, not shallower.

Conversely, the shallower profile in the direction of forward scattering is consistent with a thin disk with an $a$-dependent SPF.  To illustrate this, we used our Monte Carlo code to produce a simple circularly symmetric model of a thin disk with cross-section density $\propto r^{-2}$.  We assigned each particle in the model a HG SPF with $g = 0.9 (r/158\, {\rm AU})$.  In the region of $105 < r < 131$ AU, we found $f \propto r^{-5.6}$ along the ansae while $f \propto r^{-5.0}$ along the line of forward scattering.

We conclude that the \hd\, debris disk scale height is not sufficiently large to produce any of the observed asymmetries or trends in the SPF.  We therefore consider the \hd\, disk to be ``thin."

\subsection{Possible warping of the \hd\, debris disk: ISM interactions}

A number of debris disks exhibit clear signatures of ISM interactions \citep[e.g.][]{2007ApJ...671L.165H, 2009ApJ...702..318D}.  Interactions with the ISM gas can force dust grains out of the disk midplane, warping an otherwise thin, flat disk into three dimensions.  Our deprojection assumed the disk was thin and flat, such that the true scattering angle was constant along a given radial line to the star.  Warping by ISM gas could effectively cause the scattering angle to change with circumstellar distance, creating projection and scattering effects that were not previously considered.

To test whether ISM gas could cause the asymmetries detected in the STIS observations of the \hd\, debris disk, as well as the observed trend in the empirical scattering phase function, we produced simple models of debris disks interacting with ISM gas.  We considered only grains below the blow-out size.  Assuming the best fit dust grain composition determined by \citet{2012A&A...539A..17L}, moderate ISM interactions should result in grains below the blow-out size dominating the STIS observations of the \hd\, debris disk.  We calculate that in the absence of ISM interactions, blowout grains contribute $\sim10\%$ of the optical depth in the parent ring over the STIS bandpass, in spite of their short life times.  Barely-bound grains, predicted to dominate the parent ring cross section in the absence of ISM interactions \citep{2006ApJ...648..652S} should be easily entrained by the ISM at apastron and removed from the system quickly, resulting in the dominance of blowout grains.

We launched $200\mathord{,}000$ dust grains from a 10 AU-wide uniform parent ring centered at 90 AU.  We modeled 30 dust grain sizes below the blowout size, spaced logarithmically from $0.1$ to $4.9$ $\mu$m.  We used the best fit dust grain composition obtained by \citet{2012A&A...539A..17L} and converted grain size to $\beta$ using Figure 12 in \citet{2012A&A...539A..17L}.   We integrated the equations of motion described by \citet{2009ApJ...702..318D} using a fourth order Runge-Kutta method and a time step size equal to $0.05\%$ the orbital period at the birth ring.  We integrated until the stellocentric distances of all grains exceeded 10 times their initial stellocentric distances.  We considered both prograde and retrograde orientations for the orbits.

\hd\, has a proper motion of $23.99$ mas yr$^{-1}$ in RA and $-81.82$ mas yr$^{-1}$ in declination \citep{2007A&A...474..653V} and has a negligible radial velocity \citep{2006AstL...32..759G}, implying a velocity vector in the plane of the sky of 21 km s$^{-1}$ oriented at $16.3^{\circ}$ E of S given a distance of $51.8$ pc.  We examined 154 relative velocity vectors $\vec{v}_{\rm rel} = \vec{v}_{\star} - \vec{v}_{\rm ISM}$ between \hd\, and the ISM gas, covering 7 values of speed from 1 to 100 km s$^{-1}$ and 22 directions.  We set the density of the ISM gas to $1.67\times10^{-22}$ g cm$^{-3}$, the value used by \citet{2009ApJ...702..318D}, and used a stellar mass of $1.36$ $M_{\sun}$ \citep{2012A&A...539A..17L}.

We used \emph{dustmap} to produce images of the disk \citep{2011AJ....142..123S} using the size distribution and $Q_{\rm sca}$ values obtained by Mie theory for the best fit model of \citet{2012A&A...539A..17L}.  We assumed a single Henyey-Greenstein SPF valid for all grain sizes.  We examined 7 values of the SPF asymmetry parameter $g$ ranging from $0.2$ to $0.9$.  We note that while the observed SPF for \hd\, does not resemble a Henyey-Greenstein SPF, we chose to use the simple function for numerical rapidity; we do not intend to find a quantitative best fit to the observations.  We then reduced the modeled data using the same methods described above and searched for asymmetries qualitatively similar to those detected in the STIS image.

\begin{figure}[H]
\begin{center}
\includegraphics[width=6in]{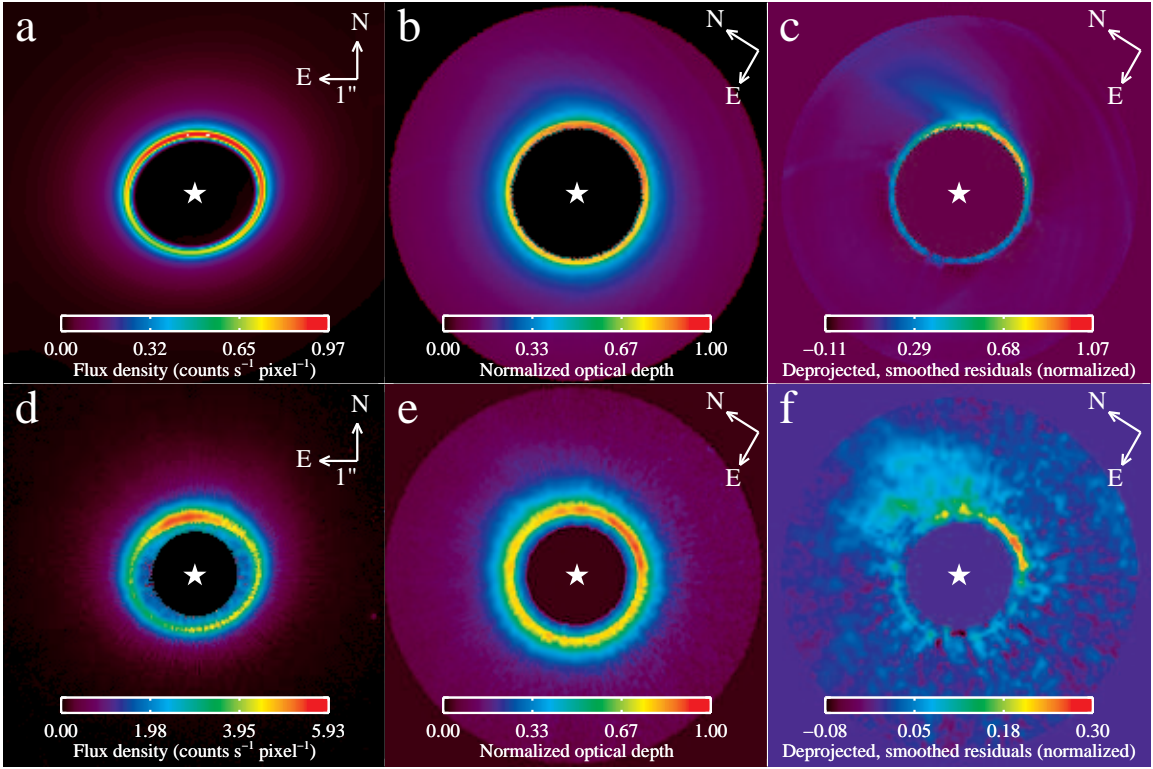}
\caption{Our best ISM-perturbed disk model (top row) compared to the STIS observations (bottom row), reduced with identical pipelines. a) Model image. b) Deprojected face-on optical depth of model.  c) Smoothed residuals of the model in the deprojected face-on plane. d) HD 181327 STIS observations. e) Deprojected face-on optical depth of STIS observations. f) Smoothed residuals of the STIS observations in the deprojected face-on plane.  The stellar location is marked in each image with a white star. \label{ISM_model_fig}}
\end{center}
\end{figure}

The top row of Figure \ref{ISM_model_fig} shows an example model in which $v_{\rm rel} = 50$ km s$^{-1}$.  We show the STIS observations of \hd\, in the bottom row for direct comparison.  The model residuals (Figure \ref{ISM_model_fig}c) appear qualitatively similar to those observed in the STIS image (Figure \ref{ISM_model_fig}f).  We are able to produce similar structures for both prograde and retrograde orbits.  We find that ISM interactions can qualitatively reproduce the bright arc along the SW portion of the parent ring if the relative velocity between \hd\, and the ISM exceeds $\sim30$ km s$^{-1}$ and is oriented toward the SW.  

Given a minimum relative velocity $v_{\rm rel} \sim 30$ km s$^{-1}$ oriented toward the SW, and a stellar velocity of $21$ km s$^{-1}$ at $16.3^{\circ}$ E of S for \hd, the ISM velocity must contribute strongly to the total relative velocity.  Roughly speaking, $v_{\rm ISM}$ must have an eastward component $\gtrsim 25$ km s$^{-1}$.  This is in contrast to other ISM-sculpted debris disks that exhibit geometries generally consistent with $v_{\rm rel}$ being dominated by the stellar velocity.

Strong ISM interactions, which should deplete barely-bound grains and increase the relative contribution of blow-out grains, may help explain the sub-micron minimum grain size inferred from fits to the SED \citep{2012A&A...539A..17L}.  Additionally, sub-micron blow-out grains, which should dominate the STIS observations, may help explain the observed degree of forward scattering, as illustrated in Figure \ref{lebreton_spf_small_grains}.

\begin{figure}[H]
\begin{center}
\includegraphics[width=4in]{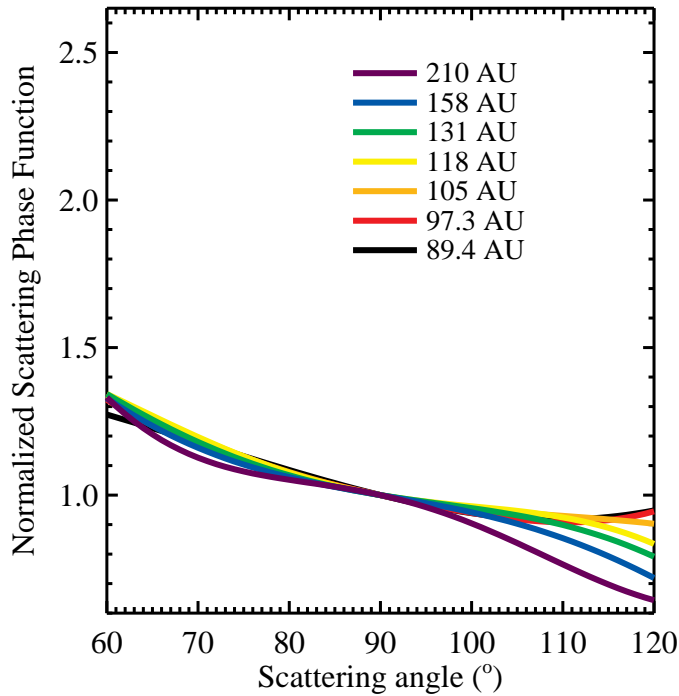}
\caption{Normalized SPF for an example ISM-perturbed disk model with $v_{\rm ISM} = 40$ km s$^{-1}$ in the direction of $37^{\circ}$ E of N and $g = 0.4$.  Our ISM-perturbed disk models can qualitatively reproduce the correct trend in the SPF as a function of $a$ for scattering angles $>90^{\circ}$, but cannot produce the correct trend for scattering angles $<90^{\circ}$ (cf. Figure \ref{spf_vs_a_figure}). \label{ISM_model_spf_vs_a_fig}}
\end{center}
\end{figure}

However, our ISM-perturbed disk models have a number of caveats.  First, blow-out grains appear to be unable to explain the trends in the SPF as a function of semi-major axis shown in Figure \ref{spf_vs_a_figure}.  The model shown in Figure \ref{ISM_model_fig} was imaged using a HG SPF with $g=0.4$ for all dust grains, independent of grain size.  Figure \ref{ISM_model_spf_vs_a_fig} shows the empirical scattering phase function derived using the techniques described in Section \ref{spf_section}.  The warping of the disk due to ISM interactions qualitatively leads to the right semi-major axis trend in the SPF for scattering angles $>90^{\circ}$, but the models cannot reproduce the correct semi-major axis trend in the SPF at scattering angles $<90^{\circ}$ (cf. Figure \ref{spf_vs_a_figure}).  We also investigated more complex scattering phase functions, including a size-dependent scattering phase function calculated from Mie theory and the empirical SPF measured in the birth ring of the STIS observations, but were unable to reproduce the trends shown in Figure \ref{spf_vs_a_figure}.

Second, the majority of ISM-perturbed disks show clear signs of a bow shock.  If $\vec{v}_{\rm rel}$ is indeed oriented toward the SW, then the \hd\, disk should show signs of a bow shock in the SW quadrant of the STIS observations; we see no signs of such a feature out to circumstellar distances $\sim 500$ AU.

Finally, to produce the asymmetry shown in Figure \ref{ISM_model_fig}, we assumed an ISM density $\sim 100$ H cm$^{-3}$, roughly three orders of magnitude greater than observed in the local bubble \citep{2009SSRv..146..235F}, where \hd\, resides \citep{2014A&A...561A..91L}.  On the other hand, the 35 pc-distant HD 61005 shows clear signs of what is thought to be ISM interaction \citep{2007ApJ...671L.165H}, and does not appear to be near any localized enhancements in ISM opacity within the local bubble \citep{2014A&A...561A..91L}.  It is unclear why we observe signs of ISM interactions in disks within the local bubble; our dynamical understanding of dust undergoing ISM drag may need revision.  Nonetheless, the large ISM density values required to produce asymmetries in this work are similar to those used for other nearby ISM-perturbed debris disks \citep{2009ApJ...702..318D}.

\section{Conclusions}
\label{conclusions}

We have imaged the \hd\ debris disk with STIS using 6-roll PSF-template subtracted coronagraphy and processed it with a new multi-roll residual removal routine (MRRR) to further reduce quasi-static PSF residuals.  The STIS observations reveal the \hd\, debris ring in its entirety.  The debris ring has a sharp inner edge and extended outer halo, consistent with a parent belt of planetesimals collisionally producing dust, and features a prominent azimuthal asymmetry.  Using a new iterative deprojection procedure, we find the disk is inclined by $28.5\pm2^{\circ}$ from face-on.  The parent ring density profile peaks at $90.5$ AU from the star assuming a distance to \hd\, of $51.8$ pc, and is nearly circular ($e=0.02\pm0.01$) with periastron located in the SW quadrant.

The empirical scattering phase function of the disk is non-Henyey-Greenstein and, due to a dramatic rise near the smallest observable scattering angles, appears significantly more forward scattering than previously thought, helping to explain the disk's low albedo.  The empirical scattering phase function also varies with semi-major axis, with more distance stellocentric distances exhibiting a greater degree of forward scattering over the range of observed scattering angles.  The scale height of the \hd\ debris disk $H/r < 0.11$, such that line-of-sight effects due to the disk thickness are negligible, and is insufficient to explain the observed trends in the scattering phase function.  If the true scattering phase function varies with semi-major axis as suggested by empirical fits, then the true radial profile of this disk's optical depth, and possibly others' measured to date, is highly uncertain.

Assuming a flat disk, we deprojected the \hd\ debris disk and removed the empirical scattering phase function to reveal the minimally-asymmetric face-on optical depth.  We find remaining asymmetries in the face-on optical depth.  The morphology of these asymmetries can be explained either by a flat disk with density enhancements due to a massive collisional event, or qualitatively by a disk warped by strong ISM interactions.

If the disk is flat and the asymmetry is due to a collisional event, the collisional mass must be greater than $10^{20}$ kg, or 1\% the mass of Pluto.  The observed trends in the scattering phase function beyond the birth ring are consistent with the radial grain size sorting predicted by models.  However, in contradiction to unperturbed debris ring models, the scattering phase function in the birth ring suggests large grains dominate the optical depth, possibly due to perturbations from an undetected planetary companion exterior to the debris ring.

If the disk is warped by the ISM, our preliminary ISM-perturbed disk models suggest the relative velocity between the ISM and star must be oriented toward the SW, at nearly a right angle to the observed stellar velocity.  This relative velocity requires an extremely dense ISM with an eastward velocity component $\gtrsim25$ km s$^{-1}$.  Our ISM-perturbed disk models cannot explain the observed changes in the degree of forward scattering as a function of semi-major axis and there are no signs of a bow shock, as is observed in other disks purportedly undergoing strong ISM interactions.

\acknowledgments

Based on observations made with the NASA/ESA Hubble Space Telescope, from program \#12228.  Support for program \#12228 was provided by NASA through a grant from the Space Telescope Science Institute, which is operated by the Association of Universities for Research in Astronomy, Inc., under NASA contract NAS 5-26555.  CCS acknowledges the support of a Carnegie Fellowship and an appointment to the NASA Postdoctoral Program at NASA Goddard Space Flight Center, administered by Oak Ridge Associated Universities through a contract with NASA.  This work was also supported by NASA Astrobiology Institute grant NNA09DA81A.

\bibliography{ms_v4}

\end{document}